\documentclass[%
 aip,
 amsmath,amssymb,
 reprint,
]{revtex4-1}

\usepackage{graphicx}
\usepackage{dcolumn}
\usepackage{bm}

\usepackage[utf8]{inputenc}
\usepackage[T1]{fontenc}
\usepackage{mathptmx}
\usepackage{etoolbox}

\usepackage{hyperref}

\draft 

\makeatletter
\def\@email#1#2{%
 \endgroup
 \patchcmd{\titleblock@produce}
  {\frontmatter@RRAPformat}
  {\frontmatter@RRAPformat{\produce@RRAP{*#1\href{mailto:#2}{#2}}}\frontmatter@RRAPformat}
  {}{}
}%
\makeatother
\begin{document}


\title{High-$Q$ membrane resonators using ultra-high-stress crystalline TiN films} 



\author{Yuki Matsuyama}
\affiliation{Komaba Institute for Science (KIS), The University of Tokyo, Meguro-ku, Tokyo 153-8902, Japan}

\author{Shotaro Shirai}
\affiliation{RIKEN Center for Quantum Computing (RQC), Wako, Saitama 351–0198, Japan}
\affiliation{Komaba Institute for Science (KIS), The University of Tokyo, Meguro-ku, Tokyo 153-8902, Japan}

\author{Ippei Nakamura}
\affiliation{Komaba Institute for Science (KIS), The University of Tokyo, Meguro-ku, Tokyo 153-8902, Japan}

\author{Masao Tokunari}
\affiliation{IBM Quantum, IBM Research - Tokyo,
Chuo-ku, Tokyo, 103-8510, Japan}

\author{Hirotaka Terai}
\affiliation{Advanced ICT Research
Institute, National Institute of Information and Communications Technology,
Kobe, Hyogo, 651-2492, Japan}

\author{Yuji Hishida}
\affiliation{Advanced ICT Research
Institute, National Institute of Information and Communications Technology,
Kobe, Hyogo, 651-2492, Japan}

\author{Ryo Sasaki}
\affiliation{RIKEN Center for Quantum Computing (RQC), Wako, Saitama 351–0198, Japan}

\author{Yusuke Tominaga}
\affiliation{RIKEN Center for Quantum Computing (RQC), Wako, Saitama 351–0198, Japan}

\author{Atsushi Noguchi}
\affiliation{Komaba Institute for Science (KIS), The University of Tokyo, Meguro-ku, Tokyo 153-8902, Japan}
\affiliation{RIKEN Center for Quantum Computing (RQC), Wako, Saitama 351–0198, Japan}
\affiliation{Inamori Research Institute for Science (InaRIS), Kyoto-shi, Kyoto 600-8411, Japan}


\date{\today}

\begin{abstract}
High-quality-factor ($Q$) mechanical resonators are essential components for precise sensing and control of mechanical motion at a quantum level.
While amorphous materials such as SiN have been widely used in high-$Q$ mechanical resonators utilizing stress-induced dissipation dilution, crystalline materials have emerging potential to achieve higher quality factors by combining low intrinsic loss and high tensile stress.
In this paper, we demonstrate high-$Q$ membrane resonators using ultra-high-stress crystalline TiN. 
Our membrane resonator exhibits a tensile stress exceeding 2.3 GPa and a quality factor of $Q = 8.0\times 10^6$ at 2.2 K.
By estimating the dilution factor, we infer that our TiN resonator has a intrinsic quality factor comparable to that of SiN membrane resonators.
With its ultra-high stress and crystalline properties, our TiN films can serve as a powerful tool for opto- and electromechanical systems, offering highly dissipation-diluted mechanical resonators.
\end{abstract}


\maketitle 
Opto- and electromechanics\cite{aspelmeyerCavityOptomechanics2014} provide a powerful platform for precise sensing and control of mechanical motion, enabling manipulation and measurement of macroscopic objects even at a quantum level.
Recent advances have demonstrated quantum control techniques such as ground-state cooling\cite{oconnellQuantumGroundState2010,teufelSidebandCoolingMicromechanical2011}, squeezed-state generation\cite{Youssefi_Kono_Chegnizadeh_Kippenberg_2023}, entanglement generation\cite{kotlerDirectObservationDeterministic2021}, and microwave-to-optical conversion.\cite{andrewsBidirectionalEfficientConversion2014,delaneySuperconductingqubitReadoutLowbackaction2022}
These developments have spurred growing interest in a broad range of applications, from quantum information technology\cite{andrewsBidirectionalEfficientConversion2014,delaneySuperconductingqubitReadoutLowbackaction2022} and quantum sensing in the search for dark matter\cite{carneyMechanicalQuantumSensing2021}, to investigation of macroscopic quantum phenomena\cite{chenMacroscopicQuantumMechanics2013}.
Since a high quality factor corresponds to long coherence and large cooperativity, pursuing high-$Q$ mechanical resonators is essential for investigating the coherent interaction between mechanical motion and electromagnetic fields.

One of the key factors for achieving high quality factor with tensioned mechanical resonators, such as membranes and beams, is the use of high tensile stress.
Mechanical resonators fabricated from highly stressed materials are known to exhibit quality factors that far exceed those expected from intrinsic material losses.
This phenomenon is known as dissipation dilution\cite{fedorovGeneralizedDissipationDilution2019,engelsenUltrahighqualityfactorMicroNanomechanical2024}.
Silicon nitride (SiN) is one of the most commonly used materials for dissipation-diluted mechanical resonators because of its high tensile stress of approximately 1 GPa\cite{southworthStressSiliconNitride2009,villanuevaEvidenceSurfaceLoss2014}, ease of fabrication, and low optical absorption\cite{zwicklHighQualityMechanical2008}.
Mechanical resonators using highly stressed SiN membranes and beams have demonstrated exceptionally high quality factors\cite{zwicklHighQualityMechanical2008,southworthStressSiliconNitride2009,schmidDampingMechanismsHigh2011,chakramDissipationUltrahighQuality2014,villanuevaEvidenceSurfaceLoss2014,yuanSiliconNitrideMembrane2015,reinhardtUltralowNoiseSiNTrampoline2016a,tsaturyanUltracoherentNanomechanicalResonators2017a,seisGroundStateCooling2022}.
Although most progress in dissipation-diluted mechanical resonators has been made using SiN, dissipation dilution can, in principle, be applicable to any material under stress\cite{fedorovGeneralizedDissipationDilution2019}.

Crystalline materials have recently emerged as promising candidates for dissipation-diluted mechanical resonators.
Films of crystalline materials can attain high tensile stress due to lattice mismatch between the film and the substrate\cite{coleTensilestrainedInxGa1XP2014,buckleStressControlTensilestrained2018}.
In addition, crystalline materials have lower defect densities than amorphous materials, allowing lower intrinsic losses, especially at cryogenic temperatures\cite{taoSinglecrystalDiamondNanomechanical2014,maccabeNanoacousticResonatorUltralong2020,engelsenUltrahighqualityfactorMicroNanomechanical2024}.
Since the quality factor of a mechanical resonator is determined by the product of the dilution factor, which is governed by tensile stress, and the intrinsic quality factor, crystalline materials have the potential to achieve extremely high quality factors.
Recently, high-$Q$ mechanical resonators have been realized using crystalline materials such as AlN\cite{ciersNanomechanicalCrystallineAlN2024}, InGaP\cite{manjeshwarHighQTrampolineResonators2023}, Si\cite{beccariStrainedCrystallineNanomechanical2022} and SiC\cite{romeroEngineeringDissipationCrystalline2020} by leveraging dissipation dilution and low intrinsic losses.
However, none of them can be employed for electromechanical systems, which requires electrodes on their structure to interact with electromagnetic fields.

Here, we present a membrane resonator using ultra-high-stress crystalline titanium nitride (TiN) films.
Since TiN is an electrically conductive material, our membrane resonator can be used in electromechanical systems without the deposition of additional metal layers, which degrade the quality factor\cite{yuControlMaterialDamping2012,seisGroundStateCooling2022} and distort the mode shape\cite{noguchiGroundStateCooling2016}.
The measured tensile stress of the TiN film on the Si substrate is more than twice that typically observed in SiN. 
We fabricated a membrane resonator incorporating a phononic crystal to suppress acoustic-radiation loss and characterized its mechanical properties via optical interferometry. 
Our membrane resonator exhibited a quality factor of $Q = 8.0\times 10^6$ for a fundamental mode at 2.2 K.
We also estimated the intrinsic quality factor of the TiN membrane resonator and found it to be comparable to that of SiN membrane resonators.


\begin{figure}
\includegraphics[width=1\linewidth]{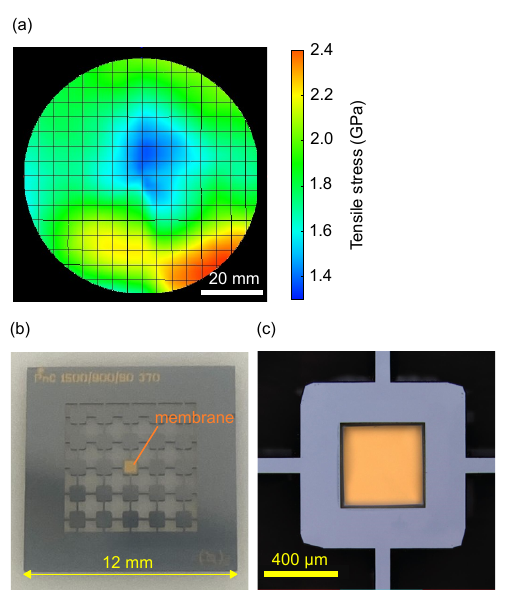}
\caption{A membrane resonator using ultra-high stressed TiN films.(a) A stress map of the TiN film on a Si substrate at room temperature. (b) A photograph of the TiN membrane resonator device.(c) A photograph of the TiN membrane taken from the backside of (b). The side length of the membrane resonator is 420$\mathrm{\mu m}$.}
\label{fig1}
\end{figure}

The membrane resonators were fabricated from a 100-nm-thick, (200)-oriented TiN film epitaxially grown at 880 $^\circ$C on a 300-$\mathrm{\mu m}$-thick silicon (100) substrate by DC magnetron sputtering\cite{sunFabrication200Oriented2015}. 
Our TiN films have already enabled high-performance superconducting qubits\cite{shiraiAllMicrowaveManipulationSuperconducting2023} and planar microwave resonators\cite{tominagaIntrinsicQualityFactors2025}, demonstrating the low intrinsic loss of the material.
In addition to the lattice mismatch between TiN and Si, the mismatch in thermal expansion coefficients between TiN and Si, and the high deposition temperature lead to the ultra-high stressed TiN films. 
A stress map of a TiN film on a 3-inch Si wafer is shown in Fig. \ref{fig1}(a).
The tensile stress was estimated from the curvature of the substrates.
We found that the tensile stress was about 1.8 GPa, which is almost twice that of typical SiN films\cite{southworthStressSiliconNitride2009,yuControlMaterialDamping2012, yuanSiliconNitrideMembrane2015}. 


We fabricated the TiN membrane resonator device shown in Fig. \ref{fig1}(b).
Our device consists of a suspended TiN membrane resonator and Si periodic structures that serve as a phononic crystal\cite{tsaturyanDemonstrationSuppressedPhonon2014,yuPhononicBandgapShield2014} (Fig. \ref{fig1}(b) and (c)). 
The purpose of the phononic crystal is to isolate the membrane modes from loss channels such as substrate modes, thus improving the quality factor and reducing the quality factor variations depending on the sample mounting configuration.
We designed the phononic crystal that has a phononic bandgap centered around the fundamental mode frequency of the membrane resonator using the finite-element-method simulation with COMSOL\cite{comsol_hp}.
The suspended membrane and phononic crystal were fabricated using anisotropic wet etching of Si crystal following the deep reactive ion etching.
For detailed fabrication processes, see the Supplementary Material.

\begin{figure}
\includegraphics[width=1\linewidth]{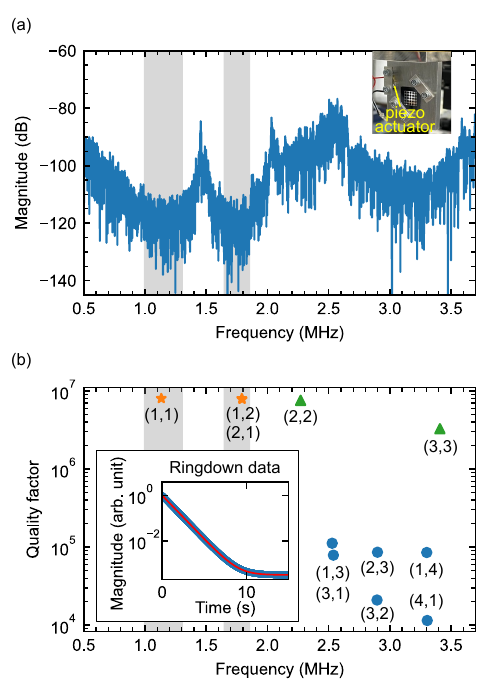}
\caption{Characterization of an ultra-high-stress TiN membrane resonator at 2.2 K. (a) Frequency response of the membrane resonator recorded at 2.2 K. Gray shaded areas indicate phononic bandgaps. Inset shows an aluminum sample mount to fix the resonator and piezo actuator. The peaks of the membrane modes are not shown due to their extremely narrow linewidth. (b) Quality factors of the membrane modes up to the (3,3) mode measured at 2.2 K. Modes inside phononic bandgaps (shaded) are shown by orange stars. Modes outside bandgaps are shown by blue circles. High-$Q$ ($>10^6$) modes outside bandgaps are shown by green triangles. Inset shows ringdown data of the fundamental mode of the membrane resonator. Time constant $\tau$ is obtained by fitting. The extracted quality factor is $8.0\times10^6$.}
\label{fig2}
\end{figure}

The mechanical properties of the TiN membrane resonator were characterized by an optical Michelson interferometer. 
The membrane resonator was mounted on an aluminum sample holder in a cryostat, whose base temperature was 2.2 K.
A piezoelectric actuator was attached next to the membrane chip to excite the membrane resonator(Fig. \ref{fig2}(a), inset).
Figure \ref{fig2} (a) shows the membrane response as a function of the drive frequency.
The signal is suppressed around 1.0--1.3 MHz and 1.65--1.85 MHz.
This is because the phononic crystal isolates the membrane resonator from non-membrane modes whose frequencies are within bandgaps.
The resonance frequency of the fundamental mode is 1.008 MHz at room temperature and 1.132 MHz at 2.2 K, both of which are within the phononic bandgap.
We note that the linewidth of each membrane mode is too narrow to be resolved in Fig. \ref{fig2} (a).

To evaluate the quality factors of the membrane modes, ringdown measurements were performed.
Fig. \ref{fig2}(b) shows the quality factors of the $(m,n)$ modes, where $m$ and $n$ are the mode indices representing the number of antinodes.
The highest quality factor $Q = 8.0\times10^6$ is recorded for the (1,1) mode.
The modes within the phononic bandgaps - (1,1), (1,2), and (2,1) - exhibit quality factors that exceed those of most modes outside the bandgaps by more than two orders of magnitude.
Despite being located outside the bandgaps, the (2,2) mode at 2.3 MHz and the (3,3) mode at 3.4 MHz exhibit high quality factors exceeding $10^6$.
This indicates that the mode shape of the membrane resonators plays a critical role in radiation loss.
For square membrane resonators, radiation loss is known to be significantly suppressed in $(n,n)$ modes\cite{wilson-raeHighNanomechanicsDestructive2011}, which is consistent with our results.

\begin{figure}
    \centering
    \includegraphics[width=\linewidth]{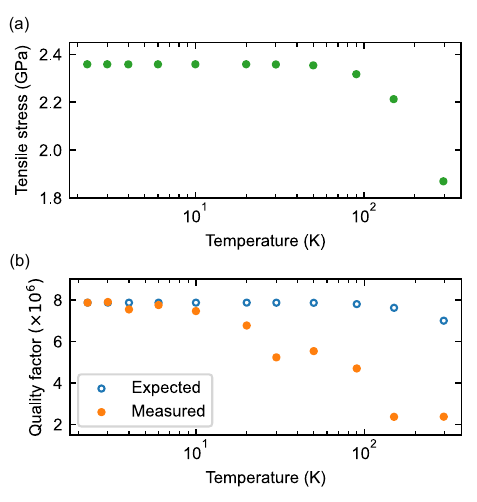}
    \caption{Temperature dependence of the mechanical properties of the membrane resonator. (a) Temperature dependence of the tensile stress on the membrane. The tensile stress is extracted from the frequency of the fundamental mode using Eq. \ref{eq:1}. (b) Temperature dependence of the quality factor of the fundamental mode. Measurement data are shown with orange filled circles. The expected reduction in quality factor with increasing temperature from 2.2 K is displayed with blue open circles, assuming only the tensile stress in Eq. \ref{eq:3} has the temperature dependence.}
    \label{fig3}
\end{figure}

We measured tensile stress and quality factors as a function of the sample stage temperature.
The temperature was varied from the base temperature up to 150 K by heating the sample stage using a temperature controller while keeping the cryostat operation.
The tensile stress was extracted from the frequency of the fundamental mode. 
Eigenfrequencies of the $(m,n)$-mode of the square membrane resonator $f_{m,n}$  are described as
\begin{equation}
   f_{m,n} = \frac{1}{2L}\sqrt{\frac{\sigma}{\rho}(m^2+n^2)},
   \label{eq:1}
\end{equation}
where $L$ is the side length of the membrane, $\sigma$ is the tensile stress, $\rho$ is the density of the material, and $m(n)$ is the numbers of antinodes for two lateral directions.
The stress was extracted under the assumption that all other parameters remain constant across the temperature range.
Figure \ref{fig3}(a) shows the temperature dependence of the stress.
The value of the tensile stress at room temperature is consistent with the measurement result shown in Fig.\ref{fig1}.
The highest value of the stress exceeds 2.3 GPa below 100 K.
The temperature dependence of the quality factor is shown in Fig. \ref{fig3}(b).
To evaluate the temperature dependence of quality factors, ringdown measurements for the fundamental mode were performed at each temperature.
The quality factor decreases with increasing the temperature and the value at room temperature is $Q = 2.4\times 10^6$.
Notably, while the tensile stress remains almost constant with increasing temperature until 50 K, the quality factor starts to decrease from 10 K.
This suggests that factors other than stress also play a role in improving the quality factor at low temperatures.
We note that the quality factor at the base temperature remained almost unchanged over four thermal cycles.

To better understand how tensile stress contributes to the observed temperature dependence of the quality factor, we considered the dissipation dilution formula\cite{yuControlMaterialDamping2012,tsaturyanUltracoherentNanomechanicalResonators2017a}.
The quality factor $Q_{m,n}$ for $(m,n)$ mode of the dissipation-diluted membrane resonator is given by
\begin{align}
    Q_{m,n} 
    &= Q_{\mathrm{int}}\times D_Q \\
    &= Q_{\mathrm{int}}\times \left(\lambda + \lambda^2\frac{(m^2+n^2)}{4}\pi^2\right)^{-1}, \label{eq:3}
\end{align}
where $Q_{\mathrm{int}}$ is the intrinsic quality factor and $D_Q$ is the dilution factor. 
$\lambda$ is given by $\lambda = \sqrt{h^2E/3L^2\sigma(1-\nu^2)}$, where $h$ is the thickness, $E$ is the Young's modulus, and $\nu$ is the Poisson's ratio.
The first term in the dilution factor corresponds to the loss near the clamping points, and the second term is associated with the loss near the antinodes.
Assuming a Young’s modulus of 200 GPa\cite{isseleDeterminationYoungsModulus2012} and a Poisson’s ratio of 0.25\cite{sunFiniteElementAnalysis1995} for the TiN thin films, we estimated the quality factor at 2.2 K to be $Q_{\mathrm{int}}\sim1.1\times 10^4$.
By imposing the temperature dependence solely on the tensile stress in Eq. \ref{eq:3}, we estimated its contribution to the observed degradation in the quality factor with increasing temperature.
The open circles in Fig. \ref{fig3}(b) represent the expected reduction in the quality factor starting from 2.2 K.
These results indicate that the contribution of the tensile stress is minor, and that the observed decrease in the quality factor is mainly attributed to the reduction in the intrinsic quality factor $Q_{\mathrm{int}}$.

\begin{figure}
    \includegraphics[width=1\linewidth]{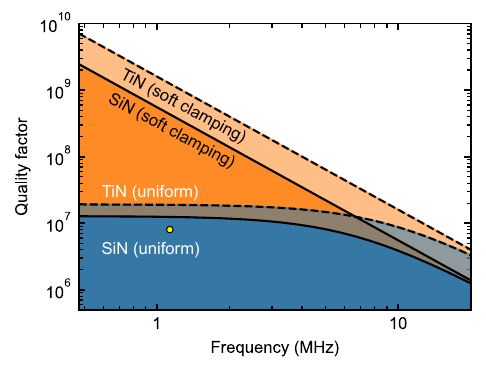}
    \caption{Comparison of achievable quality factors as a function of frequency for different materials and designs. Estimated achievable quality factors of SiN and our TiN membrane resonators with 100 nm thickness and below 1000 $\mathrm{\mu m}$ side lengths are shown as colored regions. Solid lines represent the upper limit of $Q$ for SiN, while dashed lines represent that for TiN. A yellow filled circle corresponds to the measured quality factor of the TiN membrane resonator (fundamental mode).}
    \label{fig4}
\end{figure}

Finally, we compare the performance of our TiN membrane resonator to SiN membrane resonators in previous work. 
By applying the same estimation procedure as described in the previous paragraph, the intrinsic quality factor of SiN membrane resonators shielded by phononic crystals at 10 K\cite{nielsen2017multimode} is found to be $Q_{\mathrm{int}}^{\mathrm{SiN}}\sim1.1\times 10^4$, which is comparable to that of our TiN resonators. 
Using estimated intrinsic quality factors and Eq. \ref{eq:3}, we calculate achievable quality factors of SiN and our TiN membrane resonators as a function of frequency (blue regions in Fig. \ref{fig4}). 
Although the tensile stress of TiN is approximately twice that of SiN, the achievable quality factor is only slightly higher. 
For square uniform membrane resonators, due to the extreme aspect ratio ($h/L \ll 1$), the first term of the dilution factor in Eq.\ref{eq:3} is dominant\cite{tsaturyanUltracoherentNanomechanicalResonators2017a}, and the quality factor is almost proportional to $\lambda^{-1}$
This implies that the effect of tensile stress on the quality factor scales only with the square root of the stress($Q\propto\sqrt{\sigma} $).
Therefore, under current experimental conditions, no significant advantage of TiN over SiN is observed.

To further leverage the high tensile stress of TiN to achieve higher quality factors, the soft clamping technique\cite{tsaturyanUltracoherentNanomechanicalResonators2017a} is expected to be highly effective. 
Soft clamping refers to a design that reduces energy dissipation near the clamping points of mechanical resonators.
This approach eliminates the first term of the dilution factor, resulting in a scaling of $Q \propto \lambda^{-2}$.
In other words, the quality factor becomes proportional to the tensile stress $\sigma$.
We estimate achievable quality factors for soft-clamped SiN and TiN membrane resonators (orange regions in Fig. \ref{fig4}).
By applying the soft clamping to our TiN membrane resonators, a quality factor exceeding $10^9$ can be achieved for a 1 MHz mechanical mode.

We note that our measurement temperature may not have been low enough to reveal the full potential of our crystalline TiN
membrane resonators.
It has been suggested that the temperature dependence of the dissipation of mechanical resonators is dominated by two-level systems (TLS) at low temperatures\cite{mohantyIntrinsicDissipationHighfrequency2002a,hoehneDampingHighfrequencyMetallic2010,hauerTwolevelSystemDamping2018a,zhouOnchipThermometryMicrowave2019}, regardless of whether the material is amorphous or crystalline.
According to earlier studies\cite{mohantyIntrinsicDissipationHighfrequency2002a,hoehneDampingHighfrequencyMetallic2010,hauerTwolevelSystemDamping2018a}, as the temperature decreases, mechanical dissipation tends to saturate between 10 K and 1 K, and then follows a power-law decrease below 1 K.
A notable example is a single-crystal silicon mechanical resonator\cite{maccabeNanoacousticResonatorUltralong2020}, where the quality factor improves by three orders of magnitude between 1 K and 7 mK.
Therefore, although the quality factor appears to be saturated (Fig.\ref{fig3} (b)), measurements at millikelvin temperatures will be able to demonstrate the full capability of our crystalline TiN membrane resonators.
We also note that, unlike SiN, TiN is a superconductor below 5 K, which allows TiN membrane resonators to be straightforwardly integrated with superconducting circuits including superconducting qubits.

In conclusion, we investigated the mechanical properties of the ultra-high-stress TiN membrane resonator. 
We demonstrated an ultra-high tensile stress exceeding 2.3 GPa and a high quality factor $Q = 8.0\times 10^6$ at 2.2 K.
The estimated intrinsic quality factor of our resonator at 2.2 K is comparable to that of the SiN resonator at 10 K, highlighting the potential of TiN for low-loss mechanical systems.
For membrane resonators with identical thickness, the achievable quality factor of TiN is estimated to be slightly higher than that of SiN. 
To fully exploit the high tensile stress of TiN films, the implementation of soft clamping is a promising approach.
Increasing the membrane size and reducing its thickness are also effective strategies to enhance the quality factor, as it leads to a higher dilution factor.
Future work will involve microwave measurements of TiN membrane resonators at millikelvin temperatures, as well as the exploration of device designs that further leverage the ultra-high stress of TiN films.
Such low-loss, superconducting TiN membranes may pave the way towards the realization of quantum transducers, quantum sensors, and hybrid quantum systems integrating mechanical resonators and superconducting qubits.

\vspace\baselineskip
See the supplementary material for details on device fabrication,
phononic bands calculation, and measurement setup.

\vspace\baselineskip
This work was supported by JST, CREST Grant Number JPMJCR24I5, Japan and was partly supported by IBM-UTokyo lab.

 \section*{Author declarations}
 \subsection*{Conflict of Interest}
 The authors have no conflicts to disclose.
 \subsection*{Author Contributions}
\textbf{Yuki Matsuyama}: Data Curation (lead); Formal Analysis (lead); Investigation (lead); Methodology (equal); Software (equal); Visualization (lead); Writing - Original Draft Preparation (lead)
\textbf{Shotaro Shirai}: Methodology (support); Software (equal); Writing - Review \& Editing (support)
\textbf{Ippei Nakamura}: Methodology (support); Writing - Review \& Editing (equal)
\textbf{Masao Tokunari}: Conceptualization (support); Funding Acquisition (equal); Writing - Review \& Editing (equal)
\textbf{Hirotaka Terai}: Resources (equal)
\textbf{Yuji Hishida}: Resources (equal)
\textbf{Ryo Sasaki}: Writing - Review \& Editing (equal)
\textbf{Yusuke Tominaga}: Writing - Review \& Editing (equal)
\textbf{Atsushi Noguchi}: Conceptualization (lead); Funding Acquisition (lead); Methodology (equal); Project Administration (lead); Supervision (lead); Writing - Review \& Editing (equal)

\section*{Data Availability}
The data that supports the findings of this study are available from the corresponding author upon reasonable request.



%
%

%


\bibliography{TiNmemb}

\begin{thebibliography}{44}%
\makeatletter
\providecommand \@ifxundefined [1]{%
 \@ifx{#1\undefined}
}%
\providecommand \@ifnum [1]{%
 \ifnum #1\expandafter \@firstoftwo
 \else \expandafter \@secondoftwo
 \fi
}%
\providecommand \@ifx [1]{%
 \ifx #1\expandafter \@firstoftwo
 \else \expandafter \@secondoftwo
 \fi
}%
\providecommand \natexlab [1]{#1}%
\providecommand \enquote  [1]{``#1''}%
\providecommand \bibnamefont  [1]{#1}%
\providecommand \bibfnamefont [1]{#1}%
\providecommand \citenamefont [1]{#1}%
\providecommand \href@noop [0]{\@secondoftwo}%
\providecommand \href [0]{\begingroup \@sanitize@url \@href}%
\providecommand \@href[1]{\@@startlink{#1}\@@href}%
\providecommand \@@href[1]{\endgroup#1\@@endlink}%
\providecommand \@sanitize@url [0]{\catcode `\\12\catcode `\$12\catcode `\&12\catcode `\#12\catcode `\^12\catcode `\_12\catcode `\%12\relax}%
\providecommand \@@startlink[1]{}%
\providecommand \@@endlink[0]{}%
\providecommand \url  [0]{\begingroup\@sanitize@url \@url }%
\providecommand \@url [1]{\endgroup\@href {#1}{\urlprefix }}%
\providecommand \urlprefix  [0]{URL }%
\providecommand \Eprint [0]{\href }%
\providecommand \doibase [0]{http://dx.doi.org/}%
\providecommand \selectlanguage [0]{\@gobble}%
\providecommand \bibinfo  [0]{\@secondoftwo}%
\providecommand \bibfield  [0]{\@secondoftwo}%
\providecommand \translation [1]{[#1]}%
\providecommand \BibitemOpen [0]{}%
\providecommand \bibitemStop [0]{}%
\providecommand \bibitemNoStop [0]{.\EOS\space}%
\providecommand \EOS [0]{\spacefactor3000\relax}%
\providecommand \BibitemShut  [1]{\csname bibitem#1\endcsname}%
\let\auto@bib@innerbib\@empty
\bibitem [{\citenamefont {Aspelmeyer}, \citenamefont {Kippenberg},\ and\ \citenamefont {Marquardt}(2014)}]{aspelmeyerCavityOptomechanics2014}%
  \BibitemOpen
  \bibfield  {author} {\bibinfo {author} {\bibfnamefont {M.}~\bibnamefont {Aspelmeyer}}, \bibinfo {author} {\bibfnamefont {T.~J.}\ \bibnamefont {Kippenberg}}, \ and\ \bibinfo {author} {\bibfnamefont {F.}~\bibnamefont {Marquardt}},\ }\bibfield  {title} {\enquote {\bibinfo {title} {Cavity optomechanics},}\ }\href {\doibase 10.1103/RevModPhys.86.1391} {\bibfield  {journal} {\bibinfo  {journal} {Reviews of Modern Physics}\ }\textbf {\bibinfo {volume} {86}},\ \bibinfo {pages} {1391--1452} (\bibinfo {year} {2014})}\BibitemShut {NoStop}%
\bibitem [{\citenamefont {O'Connell}\ \emph {et~al.}(2010)\citenamefont {O'Connell}, \citenamefont {Hofheinz}, \citenamefont {Ansmann}, \citenamefont {Bialczak}, \citenamefont {Lenander}, \citenamefont {Lucero}, \citenamefont {Neeley}, \citenamefont {Sank}, \citenamefont {Wang}, \citenamefont {Weides}, \citenamefont {Wenner}, \citenamefont {Martinis},\ and\ \citenamefont {Cleland}}]{oconnellQuantumGroundState2010}%
  \BibitemOpen
  \bibfield  {author} {\bibinfo {author} {\bibfnamefont {A.~D.}\ \bibnamefont {O'Connell}}, \bibinfo {author} {\bibfnamefont {M.}~\bibnamefont {Hofheinz}}, \bibinfo {author} {\bibfnamefont {M.}~\bibnamefont {Ansmann}}, \bibinfo {author} {\bibfnamefont {R.~C.}\ \bibnamefont {Bialczak}}, \bibinfo {author} {\bibfnamefont {M.}~\bibnamefont {Lenander}}, \bibinfo {author} {\bibfnamefont {E.}~\bibnamefont {Lucero}}, \bibinfo {author} {\bibfnamefont {M.}~\bibnamefont {Neeley}}, \bibinfo {author} {\bibfnamefont {D.}~\bibnamefont {Sank}}, \bibinfo {author} {\bibfnamefont {H.}~\bibnamefont {Wang}}, \bibinfo {author} {\bibfnamefont {M.}~\bibnamefont {Weides}}, \bibinfo {author} {\bibfnamefont {J.}~\bibnamefont {Wenner}}, \bibinfo {author} {\bibfnamefont {J.~M.}\ \bibnamefont {Martinis}}, \ and\ \bibinfo {author} {\bibfnamefont {A.~N.}\ \bibnamefont {Cleland}},\ }\bibfield  {title} {\enquote {\bibinfo {title} {Quantum ground state and single-phonon control of a mechanical resonator},}\ }\href {\doibase
  10.1038/nature08967} {\bibfield  {journal} {\bibinfo  {journal} {Nature}\ }\textbf {\bibinfo {volume} {464}},\ \bibinfo {pages} {697--703} (\bibinfo {year} {2010})}\BibitemShut {NoStop}%
\bibitem [{\citenamefont {Teufel}\ \emph {et~al.}(2011)\citenamefont {Teufel}, \citenamefont {Donner}, \citenamefont {Li}, \citenamefont {Harlow}, \citenamefont {Allman}, \citenamefont {Cicak}, \citenamefont {Sirois}, \citenamefont {Whittaker}, \citenamefont {Lehnert},\ and\ \citenamefont {Simmonds}}]{teufelSidebandCoolingMicromechanical2011}%
  \BibitemOpen
  \bibfield  {author} {\bibinfo {author} {\bibfnamefont {J.~D.}\ \bibnamefont {Teufel}}, \bibinfo {author} {\bibfnamefont {T.}~\bibnamefont {Donner}}, \bibinfo {author} {\bibfnamefont {D.}~\bibnamefont {Li}}, \bibinfo {author} {\bibfnamefont {J.~W.}\ \bibnamefont {Harlow}}, \bibinfo {author} {\bibfnamefont {M.~S.}\ \bibnamefont {Allman}}, \bibinfo {author} {\bibfnamefont {K.}~\bibnamefont {Cicak}}, \bibinfo {author} {\bibfnamefont {A.~J.}\ \bibnamefont {Sirois}}, \bibinfo {author} {\bibfnamefont {J.~D.}\ \bibnamefont {Whittaker}}, \bibinfo {author} {\bibfnamefont {K.~W.}\ \bibnamefont {Lehnert}}, \ and\ \bibinfo {author} {\bibfnamefont {R.~W.}\ \bibnamefont {Simmonds}},\ }\bibfield  {title} {\enquote {\bibinfo {title} {Sideband cooling of micromechanical motion to the quantum ground state},}\ }\href {\doibase 10.1038/nature10261} {\bibfield  {journal} {\bibinfo  {journal} {Nature}\ }\textbf {\bibinfo {volume} {475}},\ \bibinfo {pages} {359--363} (\bibinfo {year} {2011})}\BibitemShut {NoStop}%
\bibitem [{\citenamefont {Youssefi}\ \emph {et~al.}(2023)\citenamefont {Youssefi}, \citenamefont {Kono}, \citenamefont {Chegnizadeh},\ and\ \citenamefont {Kippenberg}}]{Youssefi_Kono_Chegnizadeh_Kippenberg_2023}%
  \BibitemOpen
  \bibfield  {author} {\bibinfo {author} {\bibfnamefont {A.}~\bibnamefont {Youssefi}}, \bibinfo {author} {\bibfnamefont {S.}~\bibnamefont {Kono}}, \bibinfo {author} {\bibfnamefont {M.}~\bibnamefont {Chegnizadeh}}, \ and\ \bibinfo {author} {\bibfnamefont {T.~J.}\ \bibnamefont {Kippenberg}},\ }\bibfield  {title} {\enquote {\bibinfo {title} {A squeezed mechanical oscillator with millisecond quantum decoherence},}\ }\href {\doibase 10.1038/s41567-023-02135-y} {\bibfield  {journal} {\bibinfo  {journal} {Nature Physics}\ }\textbf {\bibinfo {volume} {19}},\ \bibinfo {pages} {1697–1702} (\bibinfo {year} {2023})}\BibitemShut {NoStop}%
\bibitem [{\citenamefont {Kotler}\ \emph {et~al.}(2021)\citenamefont {Kotler}, \citenamefont {Peterson}, \citenamefont {Shojaee}, \citenamefont {Lecocq}, \citenamefont {Cicak}, \citenamefont {Kwiatkowski}, \citenamefont {Geller}, \citenamefont {Glancy}, \citenamefont {Knill}, \citenamefont {Simmonds}, \citenamefont {Aumentado},\ and\ \citenamefont {Teufel}}]{kotlerDirectObservationDeterministic2021}%
  \BibitemOpen
  \bibfield  {author} {\bibinfo {author} {\bibfnamefont {S.}~\bibnamefont {Kotler}}, \bibinfo {author} {\bibfnamefont {G.~A.}\ \bibnamefont {Peterson}}, \bibinfo {author} {\bibfnamefont {E.}~\bibnamefont {Shojaee}}, \bibinfo {author} {\bibfnamefont {F.}~\bibnamefont {Lecocq}}, \bibinfo {author} {\bibfnamefont {K.}~\bibnamefont {Cicak}}, \bibinfo {author} {\bibfnamefont {A.}~\bibnamefont {Kwiatkowski}}, \bibinfo {author} {\bibfnamefont {S.}~\bibnamefont {Geller}}, \bibinfo {author} {\bibfnamefont {S.}~\bibnamefont {Glancy}}, \bibinfo {author} {\bibfnamefont {E.}~\bibnamefont {Knill}}, \bibinfo {author} {\bibfnamefont {R.~W.}\ \bibnamefont {Simmonds}}, \bibinfo {author} {\bibfnamefont {J.}~\bibnamefont {Aumentado}}, \ and\ \bibinfo {author} {\bibfnamefont {J.~D.}\ \bibnamefont {Teufel}},\ }\bibfield  {title} {\enquote {\bibinfo {title} {Direct observation of deterministic macroscopic entanglement},}\ }\href {\doibase 10.1126/science.abf2998} {\bibfield  {journal} {\bibinfo  {journal} {Science}\ }\textbf
  {\bibinfo {volume} {372}},\ \bibinfo {pages} {622--625} (\bibinfo {year} {2021})}\BibitemShut {NoStop}%
\bibitem [{\citenamefont {Andrews}\ \emph {et~al.}(2014)\citenamefont {Andrews}, \citenamefont {Peterson}, \citenamefont {Purdy}, \citenamefont {Cicak}, \citenamefont {Simmonds}, \citenamefont {Regal},\ and\ \citenamefont {Lehnert}}]{andrewsBidirectionalEfficientConversion2014}%
  \BibitemOpen
  \bibfield  {author} {\bibinfo {author} {\bibfnamefont {R.~W.}\ \bibnamefont {Andrews}}, \bibinfo {author} {\bibfnamefont {R.~W.}\ \bibnamefont {Peterson}}, \bibinfo {author} {\bibfnamefont {T.~P.}\ \bibnamefont {Purdy}}, \bibinfo {author} {\bibfnamefont {K.}~\bibnamefont {Cicak}}, \bibinfo {author} {\bibfnamefont {R.~W.}\ \bibnamefont {Simmonds}}, \bibinfo {author} {\bibfnamefont {C.~A.}\ \bibnamefont {Regal}}, \ and\ \bibinfo {author} {\bibfnamefont {K.~W.}\ \bibnamefont {Lehnert}},\ }\bibfield  {title} {\enquote {\bibinfo {title} {Bidirectional and efficient conversion between microwave and optical light},}\ }\href {\doibase 10.1038/nphys2911} {\bibfield  {journal} {\bibinfo  {journal} {Nature Physics}\ }\textbf {\bibinfo {volume} {10}},\ \bibinfo {pages} {321--326} (\bibinfo {year} {2014})}\BibitemShut {NoStop}%
\bibitem [{\citenamefont {Delaney}\ \emph {et~al.}(2022)\citenamefont {Delaney}, \citenamefont {Urmey}, \citenamefont {Mittal}, \citenamefont {Brubaker}, \citenamefont {Kindem}, \citenamefont {Burns}, \citenamefont {Regal},\ and\ \citenamefont {Lehnert}}]{delaneySuperconductingqubitReadoutLowbackaction2022}%
  \BibitemOpen
  \bibfield  {author} {\bibinfo {author} {\bibfnamefont {R.~D.}\ \bibnamefont {Delaney}}, \bibinfo {author} {\bibfnamefont {M.~D.}\ \bibnamefont {Urmey}}, \bibinfo {author} {\bibfnamefont {S.}~\bibnamefont {Mittal}}, \bibinfo {author} {\bibfnamefont {B.~M.}\ \bibnamefont {Brubaker}}, \bibinfo {author} {\bibfnamefont {J.~M.}\ \bibnamefont {Kindem}}, \bibinfo {author} {\bibfnamefont {P.~S.}\ \bibnamefont {Burns}}, \bibinfo {author} {\bibfnamefont {C.~A.}\ \bibnamefont {Regal}}, \ and\ \bibinfo {author} {\bibfnamefont {K.~W.}\ \bibnamefont {Lehnert}},\ }\bibfield  {title} {\enquote {\bibinfo {title} {Superconducting-qubit readout via low-backaction electro-optic transduction},}\ }\href {\doibase 10.1038/s41586-022-04720-2} {\bibfield  {journal} {\bibinfo  {journal} {Nature}\ }\textbf {\bibinfo {volume} {606}},\ \bibinfo {pages} {489--493} (\bibinfo {year} {2022})}\BibitemShut {NoStop}%
\bibitem [{\citenamefont {Carney}\ \emph {et~al.}(2021)\citenamefont {Carney}, \citenamefont {Krnjaic}, \citenamefont {Moore}, \citenamefont {Regal}, \citenamefont {Afek}, \citenamefont {Bhave}, \citenamefont {Brubaker}, \citenamefont {Corbitt}, \citenamefont {Cripe}, \citenamefont {Crisosto}, \citenamefont {Geraci}, \citenamefont {Ghosh}, \citenamefont {Harris}, \citenamefont {Hook}, \citenamefont {Kolb}, \citenamefont {Kunjummen}, \citenamefont {Lang}, \citenamefont {Li}, \citenamefont {Lin}, \citenamefont {Liu}, \citenamefont {Lykken}, \citenamefont {Magrini}, \citenamefont {Manley}, \citenamefont {Matsumoto}, \citenamefont {Monte}, \citenamefont {Monteiro}, \citenamefont {Purdy}, \citenamefont {Riedel}, \citenamefont {Singh}, \citenamefont {Singh}, \citenamefont {Sinha}, \citenamefont {Taylor}, \citenamefont {Qin}, \citenamefont {Wilson},\ and\ \citenamefont {Zhao}}]{carneyMechanicalQuantumSensing2021}%
  \BibitemOpen
  \bibfield  {author} {\bibinfo {author} {\bibfnamefont {D.}~\bibnamefont {Carney}}, \bibinfo {author} {\bibfnamefont {G.}~\bibnamefont {Krnjaic}}, \bibinfo {author} {\bibfnamefont {D.~C.}\ \bibnamefont {Moore}}, \bibinfo {author} {\bibfnamefont {C.~A.}\ \bibnamefont {Regal}}, \bibinfo {author} {\bibfnamefont {G.}~\bibnamefont {Afek}}, \bibinfo {author} {\bibfnamefont {S.}~\bibnamefont {Bhave}}, \bibinfo {author} {\bibfnamefont {B.}~\bibnamefont {Brubaker}}, \bibinfo {author} {\bibfnamefont {T.}~\bibnamefont {Corbitt}}, \bibinfo {author} {\bibfnamefont {J.}~\bibnamefont {Cripe}}, \bibinfo {author} {\bibfnamefont {N.}~\bibnamefont {Crisosto}}, \bibinfo {author} {\bibfnamefont {A.}~\bibnamefont {Geraci}}, \bibinfo {author} {\bibfnamefont {S.}~\bibnamefont {Ghosh}}, \bibinfo {author} {\bibfnamefont {J.~G.~E.}\ \bibnamefont {Harris}}, \bibinfo {author} {\bibfnamefont {A.}~\bibnamefont {Hook}}, \bibinfo {author} {\bibfnamefont {E.~W.}\ \bibnamefont {Kolb}}, \bibinfo {author} {\bibfnamefont {J.}~\bibnamefont
  {Kunjummen}}, \bibinfo {author} {\bibfnamefont {R.~F.}\ \bibnamefont {Lang}}, \bibinfo {author} {\bibfnamefont {T.}~\bibnamefont {Li}}, \bibinfo {author} {\bibfnamefont {T.}~\bibnamefont {Lin}}, \bibinfo {author} {\bibfnamefont {Z.}~\bibnamefont {Liu}}, \bibinfo {author} {\bibfnamefont {J.}~\bibnamefont {Lykken}}, \bibinfo {author} {\bibfnamefont {L.}~\bibnamefont {Magrini}}, \bibinfo {author} {\bibfnamefont {J.}~\bibnamefont {Manley}}, \bibinfo {author} {\bibfnamefont {N.}~\bibnamefont {Matsumoto}}, \bibinfo {author} {\bibfnamefont {A.}~\bibnamefont {Monte}}, \bibinfo {author} {\bibfnamefont {F.}~\bibnamefont {Monteiro}}, \bibinfo {author} {\bibfnamefont {T.}~\bibnamefont {Purdy}}, \bibinfo {author} {\bibfnamefont {C.~J.}\ \bibnamefont {Riedel}}, \bibinfo {author} {\bibfnamefont {R.}~\bibnamefont {Singh}}, \bibinfo {author} {\bibfnamefont {S.}~\bibnamefont {Singh}}, \bibinfo {author} {\bibfnamefont {K.}~\bibnamefont {Sinha}}, \bibinfo {author} {\bibfnamefont {J.~M.}\ \bibnamefont {Taylor}}, \bibinfo
  {author} {\bibfnamefont {J.}~\bibnamefont {Qin}}, \bibinfo {author} {\bibfnamefont {D.~J.}\ \bibnamefont {Wilson}}, \ and\ \bibinfo {author} {\bibfnamefont {Y.}~\bibnamefont {Zhao}},\ }\bibfield  {title} {\enquote {\bibinfo {title} {Mechanical quantum sensing in the search for dark matter},}\ }\href {\doibase 10.1088/2058-9565/abcfcd} {\bibfield  {journal} {\bibinfo  {journal} {Quantum Science and Technology}\ }\textbf {\bibinfo {volume} {6}},\ \bibinfo {pages} {024002} (\bibinfo {year} {2021})}\BibitemShut {NoStop}%
\bibitem [{\citenamefont {Chen}(2013)}]{chenMacroscopicQuantumMechanics2013}%
  \BibitemOpen
  \bibfield  {author} {\bibinfo {author} {\bibfnamefont {Y.}~\bibnamefont {Chen}},\ }\bibfield  {title} {\enquote {\bibinfo {title} {Macroscopic quantum mechanics: Theory and experimental concepts of optomechanics},}\ }\href {\doibase 10.1088/0953-4075/46/10/104001} {\bibfield  {journal} {\bibinfo  {journal} {Journal of Physics B: Atomic, Molecular and Optical Physics}\ }\textbf {\bibinfo {volume} {46}},\ \bibinfo {pages} {104001} (\bibinfo {year} {2013})}\BibitemShut {NoStop}%
\bibitem [{\citenamefont {Fedorov}\ \emph {et~al.}(2019)\citenamefont {Fedorov}, \citenamefont {Engelsen}, \citenamefont {Ghadimi}, \citenamefont {Bereyhi}, \citenamefont {Schilling}, \citenamefont {Wilson},\ and\ \citenamefont {Kippenberg}}]{fedorovGeneralizedDissipationDilution2019}%
  \BibitemOpen
  \bibfield  {author} {\bibinfo {author} {\bibfnamefont {S.~A.}\ \bibnamefont {Fedorov}}, \bibinfo {author} {\bibfnamefont {N.~J.}\ \bibnamefont {Engelsen}}, \bibinfo {author} {\bibfnamefont {A.~H.}\ \bibnamefont {Ghadimi}}, \bibinfo {author} {\bibfnamefont {M.~J.}\ \bibnamefont {Bereyhi}}, \bibinfo {author} {\bibfnamefont {R.}~\bibnamefont {Schilling}}, \bibinfo {author} {\bibfnamefont {D.~J.}\ \bibnamefont {Wilson}}, \ and\ \bibinfo {author} {\bibfnamefont {T.~J.}\ \bibnamefont {Kippenberg}},\ }\bibfield  {title} {\enquote {\bibinfo {title} {Generalized dissipation dilution in strained mechanical resonators},}\ }\href {\doibase 10.1103/PhysRevB.99.054107} {\bibfield  {journal} {\bibinfo  {journal} {Physical Review B}\ }\textbf {\bibinfo {volume} {99}},\ \bibinfo {pages} {054107} (\bibinfo {year} {2019})}\BibitemShut {NoStop}%
\bibitem [{\citenamefont {Engelsen}, \citenamefont {Beccari},\ and\ \citenamefont {Kippenberg}(2024)}]{engelsenUltrahighqualityfactorMicroNanomechanical2024}%
  \BibitemOpen
  \bibfield  {author} {\bibinfo {author} {\bibfnamefont {N.~J.}\ \bibnamefont {Engelsen}}, \bibinfo {author} {\bibfnamefont {A.}~\bibnamefont {Beccari}}, \ and\ \bibinfo {author} {\bibfnamefont {T.~J.}\ \bibnamefont {Kippenberg}},\ }\bibfield  {title} {\enquote {\bibinfo {title} {Ultrahigh-quality-factor micro- and nanomechanical resonators using dissipation dilution},}\ }\href {\doibase 10.1038/s41565-023-01597-8} {\bibfield  {journal} {\bibinfo  {journal} {Nature Nanotechnology}\ }\textbf {\bibinfo {volume} {19}},\ \bibinfo {pages} {725--737} (\bibinfo {year} {2024})}\BibitemShut {NoStop}%
\bibitem [{\citenamefont {Southworth}\ \emph {et~al.}(2009)\citenamefont {Southworth}, \citenamefont {Barton}, \citenamefont {Verbridge}, \citenamefont {Ilic}, \citenamefont {Fefferman}, \citenamefont {Craighead},\ and\ \citenamefont {Parpia}}]{southworthStressSiliconNitride2009}%
  \BibitemOpen
  \bibfield  {author} {\bibinfo {author} {\bibfnamefont {D.~R.}\ \bibnamefont {Southworth}}, \bibinfo {author} {\bibfnamefont {R.~A.}\ \bibnamefont {Barton}}, \bibinfo {author} {\bibfnamefont {S.~S.}\ \bibnamefont {Verbridge}}, \bibinfo {author} {\bibfnamefont {B.}~\bibnamefont {Ilic}}, \bibinfo {author} {\bibfnamefont {A.~D.}\ \bibnamefont {Fefferman}}, \bibinfo {author} {\bibfnamefont {H.~G.}\ \bibnamefont {Craighead}}, \ and\ \bibinfo {author} {\bibfnamefont {J.~M.}\ \bibnamefont {Parpia}},\ }\bibfield  {title} {\enquote {\bibinfo {title} {Stress and {{Silicon Nitride}}: {{A Crack}} in the {{Universal Dissipation}} of {{Glasses}}},}\ }\href {\doibase 10.1103/PhysRevLett.102.225503} {\bibfield  {journal} {\bibinfo  {journal} {Physical Review Letters}\ }\textbf {\bibinfo {volume} {102}},\ \bibinfo {pages} {225503} (\bibinfo {year} {2009})}\BibitemShut {NoStop}%
\bibitem [{\citenamefont {Villanueva}\ and\ \citenamefont {Schmid}(2014)}]{villanuevaEvidenceSurfaceLoss2014}%
  \BibitemOpen
  \bibfield  {author} {\bibinfo {author} {\bibfnamefont {L.~G.}\ \bibnamefont {Villanueva}}\ and\ \bibinfo {author} {\bibfnamefont {S.}~\bibnamefont {Schmid}},\ }\bibfield  {title} {\enquote {\bibinfo {title} {Evidence of {{Surface Loss}} as {{Ubiquitous Limiting Damping Mechanism}} in {{SiN Micro-}} and {{Nanomechanical Resonators}}},}\ }\href {\doibase 10.1103/PhysRevLett.113.227201} {\bibfield  {journal} {\bibinfo  {journal} {Physical Review Letters}\ }\textbf {\bibinfo {volume} {113}},\ \bibinfo {pages} {227201} (\bibinfo {year} {2014})}\BibitemShut {NoStop}%
\bibitem [{\citenamefont {Zwickl}\ \emph {et~al.}(2008)\citenamefont {Zwickl}, \citenamefont {Shanks}, \citenamefont {Jayich}, \citenamefont {Yang}, \citenamefont {Bleszynski~Jayich}, \citenamefont {Thompson},\ and\ \citenamefont {Harris}}]{zwicklHighQualityMechanical2008}%
  \BibitemOpen
  \bibfield  {author} {\bibinfo {author} {\bibfnamefont {B.~M.}\ \bibnamefont {Zwickl}}, \bibinfo {author} {\bibfnamefont {W.~E.}\ \bibnamefont {Shanks}}, \bibinfo {author} {\bibfnamefont {A.~M.}\ \bibnamefont {Jayich}}, \bibinfo {author} {\bibfnamefont {C.}~\bibnamefont {Yang}}, \bibinfo {author} {\bibfnamefont {A.~C.}\ \bibnamefont {Bleszynski~Jayich}}, \bibinfo {author} {\bibfnamefont {J.~D.}\ \bibnamefont {Thompson}}, \ and\ \bibinfo {author} {\bibfnamefont {J.~G.~E.}\ \bibnamefont {Harris}},\ }\bibfield  {title} {\enquote {\bibinfo {title} {High quality mechanical and optical properties of commercial silicon nitride membranes},}\ }\href {\doibase 10.1063/1.2884191} {\bibfield  {journal} {\bibinfo  {journal} {Applied Physics Letters}\ }\textbf {\bibinfo {volume} {92}},\ \bibinfo {pages} {103125} (\bibinfo {year} {2008})}\BibitemShut {NoStop}%
\bibitem [{\citenamefont {Schmid}\ \emph {et~al.}(2011)\citenamefont {Schmid}, \citenamefont {Jensen}, \citenamefont {Nielsen},\ and\ \citenamefont {Boisen}}]{schmidDampingMechanismsHigh2011}%
  \BibitemOpen
  \bibfield  {author} {\bibinfo {author} {\bibfnamefont {S.}~\bibnamefont {Schmid}}, \bibinfo {author} {\bibfnamefont {K.~D.}\ \bibnamefont {Jensen}}, \bibinfo {author} {\bibfnamefont {K.~H.}\ \bibnamefont {Nielsen}}, \ and\ \bibinfo {author} {\bibfnamefont {A.}~\bibnamefont {Boisen}},\ }\bibfield  {title} {\enquote {\bibinfo {title} {Damping mechanisms in high- {{Q}} micro and nanomechanical string resonators},}\ }\href {\doibase 10.1103/PhysRevB.84.165307} {\bibfield  {journal} {\bibinfo  {journal} {Physical Review B}\ }\textbf {\bibinfo {volume} {84}},\ \bibinfo {pages} {165307} (\bibinfo {year} {2011})}\BibitemShut {NoStop}%
\bibitem [{\citenamefont {Chakram}\ \emph {et~al.}(2014)\citenamefont {Chakram}, \citenamefont {Patil}, \citenamefont {Chang},\ and\ \citenamefont {Vengalattore}}]{chakramDissipationUltrahighQuality2014}%
  \BibitemOpen
  \bibfield  {author} {\bibinfo {author} {\bibfnamefont {S.}~\bibnamefont {Chakram}}, \bibinfo {author} {\bibfnamefont {Y.~S.}\ \bibnamefont {Patil}}, \bibinfo {author} {\bibfnamefont {L.}~\bibnamefont {Chang}}, \ and\ \bibinfo {author} {\bibfnamefont {M.}~\bibnamefont {Vengalattore}},\ }\bibfield  {title} {\enquote {\bibinfo {title} {Dissipation in {{Ultrahigh Quality Factor SiN Membrane Resonators}}},}\ }\href {\doibase 10.1103/PhysRevLett.112.127201} {\bibfield  {journal} {\bibinfo  {journal} {Physical Review Letters}\ }\textbf {\bibinfo {volume} {112}},\ \bibinfo {pages} {127201} (\bibinfo {year} {2014})}\BibitemShut {NoStop}%
\bibitem [{\citenamefont {Yuan}, \citenamefont {Cohen},\ and\ \citenamefont {Steele}(2015)}]{yuanSiliconNitrideMembrane2015}%
  \BibitemOpen
  \bibfield  {author} {\bibinfo {author} {\bibfnamefont {M.}~\bibnamefont {Yuan}}, \bibinfo {author} {\bibfnamefont {M.~A.}\ \bibnamefont {Cohen}}, \ and\ \bibinfo {author} {\bibfnamefont {G.~A.}\ \bibnamefont {Steele}},\ }\bibfield  {title} {\enquote {\bibinfo {title} {Silicon nitride membrane resonators at millikelvin temperatures with quality factors exceeding 108},}\ }\href {\doibase 10.1063/1.4938747} {\bibfield  {journal} {\bibinfo  {journal} {Applied Physics Letters}\ }\textbf {\bibinfo {volume} {107}},\ \bibinfo {pages} {263501} (\bibinfo {year} {2015})}\BibitemShut {NoStop}%
\bibitem [{\citenamefont {Reinhardt}\ \emph {et~al.}(2016)\citenamefont {Reinhardt}, \citenamefont {M{\"u}ller}, \citenamefont {Bourassa},\ and\ \citenamefont {Sankey}}]{reinhardtUltralowNoiseSiNTrampoline2016a}%
  \BibitemOpen
  \bibfield  {author} {\bibinfo {author} {\bibfnamefont {C.}~\bibnamefont {Reinhardt}}, \bibinfo {author} {\bibfnamefont {T.}~\bibnamefont {M{\"u}ller}}, \bibinfo {author} {\bibfnamefont {A.}~\bibnamefont {Bourassa}}, \ and\ \bibinfo {author} {\bibfnamefont {J.~C.}\ \bibnamefont {Sankey}},\ }\bibfield  {title} {\enquote {\bibinfo {title} {Ultralow-{{Noise SiN Trampoline Resonators}} for {{Sensing}} and {{Optomechanics}}},}\ }\href {\doibase 10.1103/PhysRevX.6.021001} {\bibfield  {journal} {\bibinfo  {journal} {Physical Review X}\ }\textbf {\bibinfo {volume} {6}},\ \bibinfo {pages} {021001} (\bibinfo {year} {2016})}\BibitemShut {NoStop}%
\bibitem [{\citenamefont {Tsaturyan}\ \emph {et~al.}(2017)\citenamefont {Tsaturyan}, \citenamefont {Barg}, \citenamefont {Polzik},\ and\ \citenamefont {Schliesser}}]{tsaturyanUltracoherentNanomechanicalResonators2017a}%
  \BibitemOpen
  \bibfield  {author} {\bibinfo {author} {\bibfnamefont {Y.}~\bibnamefont {Tsaturyan}}, \bibinfo {author} {\bibfnamefont {A.}~\bibnamefont {Barg}}, \bibinfo {author} {\bibfnamefont {E.~S.}\ \bibnamefont {Polzik}}, \ and\ \bibinfo {author} {\bibfnamefont {A.}~\bibnamefont {Schliesser}},\ }\bibfield  {title} {\enquote {\bibinfo {title} {Ultracoherent nanomechanical resonators via soft clamping and dissipation dilution},}\ }\href {\doibase 10.1038/nnano.2017.101} {\bibfield  {journal} {\bibinfo  {journal} {Nature Nanotechnology}\ }\textbf {\bibinfo {volume} {12}},\ \bibinfo {pages} {776--783} (\bibinfo {year} {2017})}\BibitemShut {NoStop}%
\bibitem [{\citenamefont {Seis}\ \emph {et~al.}(2022)\citenamefont {Seis}, \citenamefont {Capelle}, \citenamefont {Langman}, \citenamefont {Saarinen}, \citenamefont {Planz},\ and\ \citenamefont {Schliesser}}]{seisGroundStateCooling2022}%
  \BibitemOpen
  \bibfield  {author} {\bibinfo {author} {\bibfnamefont {Y.}~\bibnamefont {Seis}}, \bibinfo {author} {\bibfnamefont {T.}~\bibnamefont {Capelle}}, \bibinfo {author} {\bibfnamefont {E.}~\bibnamefont {Langman}}, \bibinfo {author} {\bibfnamefont {S.}~\bibnamefont {Saarinen}}, \bibinfo {author} {\bibfnamefont {E.}~\bibnamefont {Planz}}, \ and\ \bibinfo {author} {\bibfnamefont {A.}~\bibnamefont {Schliesser}},\ }\bibfield  {title} {\enquote {\bibinfo {title} {Ground state cooling of an ultracoherent electromechanical system},}\ }\href {\doibase 10.1038/s41467-022-29115-9} {\bibfield  {journal} {\bibinfo  {journal} {Nature Communications}\ }\textbf {\bibinfo {volume} {13}},\ \bibinfo {pages} {1507} (\bibinfo {year} {2022})}\BibitemShut {NoStop}%
\bibitem [{\citenamefont {Cole}\ \emph {et~al.}(2014)\citenamefont {Cole}, \citenamefont {Yu}, \citenamefont {G{\"a}rtner}, \citenamefont {Siquans}, \citenamefont {Moghadas~Nia}, \citenamefont {Schm{\"o}le}, \citenamefont {{Hoelscher-Obermaier}}, \citenamefont {Purdy}, \citenamefont {Wieczorek}, \citenamefont {Regal},\ and\ \citenamefont {Aspelmeyer}}]{coleTensilestrainedInxGa1XP2014}%
  \BibitemOpen
  \bibfield  {author} {\bibinfo {author} {\bibfnamefont {G.~D.}\ \bibnamefont {Cole}}, \bibinfo {author} {\bibfnamefont {P.-L.}\ \bibnamefont {Yu}}, \bibinfo {author} {\bibfnamefont {C.}~\bibnamefont {G{\"a}rtner}}, \bibinfo {author} {\bibfnamefont {K.}~\bibnamefont {Siquans}}, \bibinfo {author} {\bibfnamefont {R.}~\bibnamefont {Moghadas~Nia}}, \bibinfo {author} {\bibfnamefont {J.}~\bibnamefont {Schm{\"o}le}}, \bibinfo {author} {\bibfnamefont {J.}~\bibnamefont {{Hoelscher-Obermaier}}}, \bibinfo {author} {\bibfnamefont {T.~P.}\ \bibnamefont {Purdy}}, \bibinfo {author} {\bibfnamefont {W.}~\bibnamefont {Wieczorek}}, \bibinfo {author} {\bibfnamefont {C.~A.}\ \bibnamefont {Regal}}, \ and\ \bibinfo {author} {\bibfnamefont {M.}~\bibnamefont {Aspelmeyer}},\ }\bibfield  {title} {\enquote {\bibinfo {title} {Tensile-strained {{InxGa1}}-{{xP}} membranes for cavity optomechanics},}\ }\href {\doibase 10.1063/1.4879755} {\bibfield  {journal} {\bibinfo  {journal} {Applied Physics Letters}\ }\textbf {\bibinfo {volume}
  {104}},\ \bibinfo {pages} {201908} (\bibinfo {year} {2014})}\BibitemShut {NoStop}%
\bibitem [{\citenamefont {B{\"u}ckle}\ \emph {et~al.}(2018)\citenamefont {B{\"u}ckle}, \citenamefont {Hauber}, \citenamefont {Cole}, \citenamefont {G{\"a}rtner}, \citenamefont {Zeimer}, \citenamefont {Grenzer},\ and\ \citenamefont {Weig}}]{buckleStressControlTensilestrained2018}%
  \BibitemOpen
  \bibfield  {author} {\bibinfo {author} {\bibfnamefont {M.}~\bibnamefont {B{\"u}ckle}}, \bibinfo {author} {\bibfnamefont {V.~C.}\ \bibnamefont {Hauber}}, \bibinfo {author} {\bibfnamefont {G.~D.}\ \bibnamefont {Cole}}, \bibinfo {author} {\bibfnamefont {C.}~\bibnamefont {G{\"a}rtner}}, \bibinfo {author} {\bibfnamefont {U.}~\bibnamefont {Zeimer}}, \bibinfo {author} {\bibfnamefont {J.}~\bibnamefont {Grenzer}}, \ and\ \bibinfo {author} {\bibfnamefont {E.~M.}\ \bibnamefont {Weig}},\ }\bibfield  {title} {\enquote {\bibinfo {title} {Stress control of tensile-strained {{In1}}- {\emph{x}} {{Ga}} {\emph{x}} {{P}} nanomechanical string resonators},}\ }\href {\doibase 10.1063/1.5054076} {\bibfield  {journal} {\bibinfo  {journal} {Applied Physics Letters}\ }\textbf {\bibinfo {volume} {113}},\ \bibinfo {pages} {201903} (\bibinfo {year} {2018})}\BibitemShut {NoStop}%
\bibitem [{\citenamefont {Tao}\ \emph {et~al.}(2014)\citenamefont {Tao}, \citenamefont {Boss}, \citenamefont {Moores},\ and\ \citenamefont {Degen}}]{taoSinglecrystalDiamondNanomechanical2014}%
  \BibitemOpen
  \bibfield  {author} {\bibinfo {author} {\bibfnamefont {Y.}~\bibnamefont {Tao}}, \bibinfo {author} {\bibfnamefont {J.~M.}\ \bibnamefont {Boss}}, \bibinfo {author} {\bibfnamefont {B.~A.}\ \bibnamefont {Moores}}, \ and\ \bibinfo {author} {\bibfnamefont {C.~L.}\ \bibnamefont {Degen}},\ }\bibfield  {title} {\enquote {\bibinfo {title} {Single-crystal diamond nanomechanical resonators with quality factors exceeding one million},}\ }\href {\doibase 10.1038/ncomms4638} {\bibfield  {journal} {\bibinfo  {journal} {Nature Communications}\ }\textbf {\bibinfo {volume} {5}},\ \bibinfo {pages} {3638} (\bibinfo {year} {2014})}\BibitemShut {NoStop}%
\bibitem [{\citenamefont {MacCabe}\ \emph {et~al.}(2020)\citenamefont {MacCabe}, \citenamefont {Ren}, \citenamefont {Luo}, \citenamefont {Cohen}, \citenamefont {Zhou}, \citenamefont {Sipahigil}, \citenamefont {Mirhosseini},\ and\ \citenamefont {Painter}}]{maccabeNanoacousticResonatorUltralong2020}%
  \BibitemOpen
  \bibfield  {author} {\bibinfo {author} {\bibfnamefont {G.~S.}\ \bibnamefont {MacCabe}}, \bibinfo {author} {\bibfnamefont {H.}~\bibnamefont {Ren}}, \bibinfo {author} {\bibfnamefont {J.}~\bibnamefont {Luo}}, \bibinfo {author} {\bibfnamefont {J.~D.}\ \bibnamefont {Cohen}}, \bibinfo {author} {\bibfnamefont {H.}~\bibnamefont {Zhou}}, \bibinfo {author} {\bibfnamefont {A.}~\bibnamefont {Sipahigil}}, \bibinfo {author} {\bibfnamefont {M.}~\bibnamefont {Mirhosseini}}, \ and\ \bibinfo {author} {\bibfnamefont {O.}~\bibnamefont {Painter}},\ }\bibfield  {title} {\enquote {\bibinfo {title} {Nano-acoustic resonator with ultralong phonon lifetime},}\ }\href {\doibase 10.1126/science.abc7312} {\bibfield  {journal} {\bibinfo  {journal} {Science}\ }\textbf {\bibinfo {volume} {370}},\ \bibinfo {pages} {840--843} (\bibinfo {year} {2020})}\BibitemShut {NoStop}%
\bibitem [{\citenamefont {Ciers}\ \emph {et~al.}(2024)\citenamefont {Ciers}, \citenamefont {Jung}, \citenamefont {Ciers}, \citenamefont {Nindito}, \citenamefont {Pfeifer}, \citenamefont {Dadgar}, \citenamefont {Strittmatter},\ and\ \citenamefont {Wieczorek}}]{ciersNanomechanicalCrystallineAlN2024}%
  \BibitemOpen
  \bibfield  {author} {\bibinfo {author} {\bibfnamefont {A.}~\bibnamefont {Ciers}}, \bibinfo {author} {\bibfnamefont {A.}~\bibnamefont {Jung}}, \bibinfo {author} {\bibfnamefont {J.}~\bibnamefont {Ciers}}, \bibinfo {author} {\bibfnamefont {L.~R.}\ \bibnamefont {Nindito}}, \bibinfo {author} {\bibfnamefont {H.}~\bibnamefont {Pfeifer}}, \bibinfo {author} {\bibfnamefont {A.}~\bibnamefont {Dadgar}}, \bibinfo {author} {\bibfnamefont {A.}~\bibnamefont {Strittmatter}}, \ and\ \bibinfo {author} {\bibfnamefont {W.}~\bibnamefont {Wieczorek}},\ }\bibfield  {title} {\enquote {\bibinfo {title} {Nanomechanical {{Crystalline AlN Resonators}} with {{High Quality Factors}} for {{Quantum Optoelectromechanics}}},}\ }\href {\doibase 10.1002/adma.202403155} {\bibfield  {journal} {\bibinfo  {journal} {Advanced Materials}\ }\textbf {\bibinfo {volume} {36}},\ \bibinfo {pages} {2403155} (\bibinfo {year} {2024})}\BibitemShut {NoStop}%
\bibitem [{\citenamefont {Manjeshwar}\ \emph {et~al.}(2023)\citenamefont {Manjeshwar}, \citenamefont {Ciers}, \citenamefont {Hellman}, \citenamefont {Bl{\"a}sing}, \citenamefont {Strittmatter},\ and\ \citenamefont {Wieczorek}}]{manjeshwarHighQTrampolineResonators2023}%
  \BibitemOpen
  \bibfield  {author} {\bibinfo {author} {\bibfnamefont {S.~K.}\ \bibnamefont {Manjeshwar}}, \bibinfo {author} {\bibfnamefont {A.}~\bibnamefont {Ciers}}, \bibinfo {author} {\bibfnamefont {F.}~\bibnamefont {Hellman}}, \bibinfo {author} {\bibfnamefont {J.}~\bibnamefont {Bl{\"a}sing}}, \bibinfo {author} {\bibfnamefont {A.}~\bibnamefont {Strittmatter}}, \ and\ \bibinfo {author} {\bibfnamefont {W.}~\bibnamefont {Wieczorek}},\ }\bibfield  {title} {\enquote {\bibinfo {title} {High-{{Q Trampoline Resonators}} from {{Strained Crystalline InGaP}} for {{Integrated Free-Space Optomechanics}}},}\ }\href {\doibase 10.1021/acs.nanolett.3c00996} {\bibfield  {journal} {\bibinfo  {journal} {Nano Letters}\ }\textbf {\bibinfo {volume} {23}},\ \bibinfo {pages} {5076--5082} (\bibinfo {year} {2023})}\BibitemShut {NoStop}%
\bibitem [{\citenamefont {Beccari}\ \emph {et~al.}(2022)\citenamefont {Beccari}, \citenamefont {Visani}, \citenamefont {Fedorov}, \citenamefont {Bereyhi}, \citenamefont {Boureau}, \citenamefont {Engelsen},\ and\ \citenamefont {Kippenberg}}]{beccariStrainedCrystallineNanomechanical2022}%
  \BibitemOpen
  \bibfield  {author} {\bibinfo {author} {\bibfnamefont {A.}~\bibnamefont {Beccari}}, \bibinfo {author} {\bibfnamefont {D.~A.}\ \bibnamefont {Visani}}, \bibinfo {author} {\bibfnamefont {S.~A.}\ \bibnamefont {Fedorov}}, \bibinfo {author} {\bibfnamefont {M.~J.}\ \bibnamefont {Bereyhi}}, \bibinfo {author} {\bibfnamefont {V.}~\bibnamefont {Boureau}}, \bibinfo {author} {\bibfnamefont {N.~J.}\ \bibnamefont {Engelsen}}, \ and\ \bibinfo {author} {\bibfnamefont {T.~J.}\ \bibnamefont {Kippenberg}},\ }\bibfield  {title} {\enquote {\bibinfo {title} {Strained crystalline nanomechanical resonators with quality factors above 10 billion},}\ }\href {\doibase 10.1038/s41567-021-01498-4} {\bibfield  {journal} {\bibinfo  {journal} {Nature Physics}\ }\textbf {\bibinfo {volume} {18}},\ \bibinfo {pages} {436--441} (\bibinfo {year} {2022})}\BibitemShut {NoStop}%
\bibitem [{\citenamefont {Romero}\ \emph {et~al.}(2020)\citenamefont {Romero}, \citenamefont {Valenzuela}, \citenamefont {Kermany}, \citenamefont {Sementilli}, \citenamefont {Iacopi},\ and\ \citenamefont {Bowen}}]{romeroEngineeringDissipationCrystalline2020}%
  \BibitemOpen
  \bibfield  {author} {\bibinfo {author} {\bibfnamefont {E.}~\bibnamefont {Romero}}, \bibinfo {author} {\bibfnamefont {V.~M.}\ \bibnamefont {Valenzuela}}, \bibinfo {author} {\bibfnamefont {A.~R.}\ \bibnamefont {Kermany}}, \bibinfo {author} {\bibfnamefont {L.}~\bibnamefont {Sementilli}}, \bibinfo {author} {\bibfnamefont {F.}~\bibnamefont {Iacopi}}, \ and\ \bibinfo {author} {\bibfnamefont {W.~P.}\ \bibnamefont {Bowen}},\ }\bibfield  {title} {\enquote {\bibinfo {title} {Engineering the {{Dissipation}} of {{Crystalline Micromechanical Resonators}}},}\ }\href {\doibase 10.1103/PhysRevApplied.13.044007} {\bibfield  {journal} {\bibinfo  {journal} {Physical Review Applied}\ }\textbf {\bibinfo {volume} {13}},\ \bibinfo {pages} {044007} (\bibinfo {year} {2020})}\BibitemShut {NoStop}%
\bibitem [{\citenamefont {Yu}, \citenamefont {Purdy},\ and\ \citenamefont {Regal}(2012)}]{yuControlMaterialDamping2012}%
  \BibitemOpen
  \bibfield  {author} {\bibinfo {author} {\bibfnamefont {P.-L.}\ \bibnamefont {Yu}}, \bibinfo {author} {\bibfnamefont {T.~P.}\ \bibnamefont {Purdy}}, \ and\ \bibinfo {author} {\bibfnamefont {C.~A.}\ \bibnamefont {Regal}},\ }\bibfield  {title} {\enquote {\bibinfo {title} {Control of {{Material Damping}} in {{High- Q Membrane Microresonators}}},}\ }\href {\doibase 10.1103/PhysRevLett.108.083603} {\bibfield  {journal} {\bibinfo  {journal} {Physical Review Letters}\ }\textbf {\bibinfo {volume} {108}},\ \bibinfo {pages} {083603} (\bibinfo {year} {2012})}\BibitemShut {NoStop}%
\bibitem [{\citenamefont {Noguchi}\ \emph {et~al.}(2016)\citenamefont {Noguchi}, \citenamefont {Yamazaki}, \citenamefont {Ataka}, \citenamefont {Fujita}, \citenamefont {Tabuchi}, \citenamefont {Ishikawa}, \citenamefont {Usami},\ and\ \citenamefont {Nakamura}}]{noguchiGroundStateCooling2016}%
  \BibitemOpen
  \bibfield  {author} {\bibinfo {author} {\bibfnamefont {A.}~\bibnamefont {Noguchi}}, \bibinfo {author} {\bibfnamefont {R.}~\bibnamefont {Yamazaki}}, \bibinfo {author} {\bibfnamefont {M.}~\bibnamefont {Ataka}}, \bibinfo {author} {\bibfnamefont {H.}~\bibnamefont {Fujita}}, \bibinfo {author} {\bibfnamefont {Y.}~\bibnamefont {Tabuchi}}, \bibinfo {author} {\bibfnamefont {T.}~\bibnamefont {Ishikawa}}, \bibinfo {author} {\bibfnamefont {K.}~\bibnamefont {Usami}}, \ and\ \bibinfo {author} {\bibfnamefont {Y.}~\bibnamefont {Nakamura}},\ }\bibfield  {title} {\enquote {\bibinfo {title} {Ground state cooling of a quantum electromechanical system with a silicon nitride membrane in a {{3D}} loop-gap cavity},}\ }\href {\doibase 10.1088/1367-2630/18/10/103036} {\bibfield  {journal} {\bibinfo  {journal} {New Journal of Physics}\ }\textbf {\bibinfo {volume} {18}},\ \bibinfo {pages} {103036} (\bibinfo {year} {2016})}\BibitemShut {NoStop}%
\bibitem [{\citenamefont {Sun}\ \emph {et~al.}(2015)\citenamefont {Sun}, \citenamefont {Makise}, \citenamefont {Qiu}, \citenamefont {Terai},\ and\ \citenamefont {Wang}}]{sunFabrication200Oriented2015}%
  \BibitemOpen
  \bibfield  {author} {\bibinfo {author} {\bibfnamefont {R.}~\bibnamefont {Sun}}, \bibinfo {author} {\bibfnamefont {K.}~\bibnamefont {Makise}}, \bibinfo {author} {\bibfnamefont {W.}~\bibnamefont {Qiu}}, \bibinfo {author} {\bibfnamefont {H.}~\bibnamefont {Terai}}, \ and\ \bibinfo {author} {\bibfnamefont {Z.}~\bibnamefont {Wang}},\ }\bibfield  {title} {\enquote {\bibinfo {title} {Fabrication of (200)-{{Oriented TiN Films}} on {{Si}} (100) {{Substrates}} by {{DC Magnetron Sputtering}}},}\ }\href {\doibase 10.1109/TASC.2014.2383694} {\bibfield  {journal} {\bibinfo  {journal} {IEEE Transactions on Applied Superconductivity}\ }\textbf {\bibinfo {volume} {25}},\ \bibinfo {pages} {1--4} (\bibinfo {year} {2015})}\BibitemShut {NoStop}%
\bibitem [{\citenamefont {Shirai}\ \emph {et~al.}(2023)\citenamefont {Shirai}, \citenamefont {Okubo}, \citenamefont {Matsuura}, \citenamefont {Osada}, \citenamefont {Nakamura},\ and\ \citenamefont {Noguchi}}]{shiraiAllMicrowaveManipulationSuperconducting2023}%
  \BibitemOpen
  \bibfield  {author} {\bibinfo {author} {\bibfnamefont {S.}~\bibnamefont {Shirai}}, \bibinfo {author} {\bibfnamefont {Y.}~\bibnamefont {Okubo}}, \bibinfo {author} {\bibfnamefont {K.}~\bibnamefont {Matsuura}}, \bibinfo {author} {\bibfnamefont {A.}~\bibnamefont {Osada}}, \bibinfo {author} {\bibfnamefont {Y.}~\bibnamefont {Nakamura}}, \ and\ \bibinfo {author} {\bibfnamefont {A.}~\bibnamefont {Noguchi}},\ }\bibfield  {title} {\enquote {\bibinfo {title} {All-{{Microwave Manipulation}} of {{Superconducting Qubits}} with a {{Fixed-Frequency Transmon Coupler}}},}\ }\href {\doibase 10.1103/PhysRevLett.130.260601} {\bibfield  {journal} {\bibinfo  {journal} {Physical Review Letters}\ }\textbf {\bibinfo {volume} {130}},\ \bibinfo {pages} {260601} (\bibinfo {year} {2023})}\BibitemShut {NoStop}%
\bibitem [{\citenamefont {Tominaga}\ \emph {et~al.}(2025)\citenamefont {Tominaga}, \citenamefont {Shirai}, \citenamefont {Hishida}, \citenamefont {Terai},\ and\ \citenamefont {Noguchi}}]{tominagaIntrinsicQualityFactors2025}%
  \BibitemOpen
  \bibfield  {author} {\bibinfo {author} {\bibfnamefont {Y.}~\bibnamefont {Tominaga}}, \bibinfo {author} {\bibfnamefont {S.}~\bibnamefont {Shirai}}, \bibinfo {author} {\bibfnamefont {Y.}~\bibnamefont {Hishida}}, \bibinfo {author} {\bibfnamefont {H.}~\bibnamefont {Terai}}, \ and\ \bibinfo {author} {\bibfnamefont {A.}~\bibnamefont {Noguchi}},\ }\bibfield  {title} {\enquote {\bibinfo {title} {Intrinsic quality factors approaching 10 million in superconducting planar resonators enabled by spiral geometry},}\ }\href {\doibase 10.1140/epjqt/s40507-025-00367-w} {\bibfield  {journal} {\bibinfo  {journal} {EPJ Quantum Technology}\ }\textbf {\bibinfo {volume} {12}},\ \bibinfo {pages} {1--13} (\bibinfo {year} {2025})}\BibitemShut {NoStop}%
\bibitem [{\citenamefont {Tsaturyan}\ \emph {et~al.}(2014)\citenamefont {Tsaturyan}, \citenamefont {Barg}, \citenamefont {Simonsen}, \citenamefont {Villanueva}, \citenamefont {Schmid}, \citenamefont {Schliesser},\ and\ \citenamefont {Polzik}}]{tsaturyanDemonstrationSuppressedPhonon2014}%
  \BibitemOpen
  \bibfield  {author} {\bibinfo {author} {\bibfnamefont {Y.}~\bibnamefont {Tsaturyan}}, \bibinfo {author} {\bibfnamefont {A.}~\bibnamefont {Barg}}, \bibinfo {author} {\bibfnamefont {A.}~\bibnamefont {Simonsen}}, \bibinfo {author} {\bibfnamefont {L.~G.}\ \bibnamefont {Villanueva}}, \bibinfo {author} {\bibfnamefont {S.}~\bibnamefont {Schmid}}, \bibinfo {author} {\bibfnamefont {A.}~\bibnamefont {Schliesser}}, \ and\ \bibinfo {author} {\bibfnamefont {E.~S.}\ \bibnamefont {Polzik}},\ }\bibfield  {title} {\enquote {\bibinfo {title} {Demonstration of suppressed phonon tunneling losses in phononic bandgap shielded membrane resonators for high-{{Q}} optomechanics},}\ }\href {\doibase 10.1364/OE.22.006810} {\bibfield  {journal} {\bibinfo  {journal} {Optics Express}\ }\textbf {\bibinfo {volume} {22}},\ \bibinfo {pages} {6810} (\bibinfo {year} {2014})}\BibitemShut {NoStop}%
\bibitem [{\citenamefont {Yu}\ \emph {et~al.}(2014)\citenamefont {Yu}, \citenamefont {Cicak}, \citenamefont {Kampel}, \citenamefont {Tsaturyan}, \citenamefont {Purdy}, \citenamefont {Simmonds},\ and\ \citenamefont {Regal}}]{yuPhononicBandgapShield2014}%
  \BibitemOpen
  \bibfield  {author} {\bibinfo {author} {\bibfnamefont {P.-L.}\ \bibnamefont {Yu}}, \bibinfo {author} {\bibfnamefont {K.}~\bibnamefont {Cicak}}, \bibinfo {author} {\bibfnamefont {N.~S.}\ \bibnamefont {Kampel}}, \bibinfo {author} {\bibfnamefont {Y.}~\bibnamefont {Tsaturyan}}, \bibinfo {author} {\bibfnamefont {T.~P.}\ \bibnamefont {Purdy}}, \bibinfo {author} {\bibfnamefont {R.~W.}\ \bibnamefont {Simmonds}}, \ and\ \bibinfo {author} {\bibfnamefont {C.~A.}\ \bibnamefont {Regal}},\ }\bibfield  {title} {\enquote {\bibinfo {title} {A phononic bandgap shield for high- {{{\emph{Q}}}} membrane microresonators},}\ }\href {\doibase 10.1063/1.4862031} {\bibfield  {journal} {\bibinfo  {journal} {Applied Physics Letters}\ }\textbf {\bibinfo {volume} {104}},\ \bibinfo {pages} {023510} (\bibinfo {year} {2014})}\BibitemShut {NoStop}%
\bibitem [{com()}]{comsol_hp}%
  \BibitemOpen
  \href@noop {} {\enquote {\bibinfo {title} {Comsol 6.0},}\ }\bibinfo {howpublished} {\url{https://www.comsol.jp/}}\BibitemShut {NoStop}%
\bibitem [{\citenamefont {{Wilson-Rae}}\ \emph {et~al.}(2011)\citenamefont {{Wilson-Rae}}, \citenamefont {Barton}, \citenamefont {Verbridge}, \citenamefont {Southworth}, \citenamefont {Ilic}, \citenamefont {Craighead},\ and\ \citenamefont {Parpia}}]{wilson-raeHighNanomechanicsDestructive2011}%
  \BibitemOpen
  \bibfield  {author} {\bibinfo {author} {\bibfnamefont {I.}~\bibnamefont {{Wilson-Rae}}}, \bibinfo {author} {\bibfnamefont {R.~A.}\ \bibnamefont {Barton}}, \bibinfo {author} {\bibfnamefont {S.~S.}\ \bibnamefont {Verbridge}}, \bibinfo {author} {\bibfnamefont {D.~R.}\ \bibnamefont {Southworth}}, \bibinfo {author} {\bibfnamefont {B.}~\bibnamefont {Ilic}}, \bibinfo {author} {\bibfnamefont {H.~G.}\ \bibnamefont {Craighead}}, \ and\ \bibinfo {author} {\bibfnamefont {J.~M.}\ \bibnamefont {Parpia}},\ }\bibfield  {title} {\enquote {\bibinfo {title} {High- {{Q Nanomechanics}} via {{Destructive Interference}} of {{Elastic Waves}}},}\ }\href {\doibase 10.1103/PhysRevLett.106.047205} {\bibfield  {journal} {\bibinfo  {journal} {Physical Review Letters}\ }\textbf {\bibinfo {volume} {106}},\ \bibinfo {pages} {047205} (\bibinfo {year} {2011})}\BibitemShut {NoStop}%
\bibitem [{\citenamefont {Issel{\'e}}\ \emph {et~al.}(2012)\citenamefont {Issel{\'e}}, \citenamefont {Mercier}, \citenamefont {Parry}, \citenamefont {Estevez}, \citenamefont {Vignoud},\ and\ \citenamefont {Olagnon}}]{isseleDeterminationYoungsModulus2012}%
  \BibitemOpen
  \bibfield  {author} {\bibinfo {author} {\bibfnamefont {H.}~\bibnamefont {Issel{\'e}}}, \bibinfo {author} {\bibfnamefont {D.}~\bibnamefont {Mercier}}, \bibinfo {author} {\bibfnamefont {G.}~\bibnamefont {Parry}}, \bibinfo {author} {\bibfnamefont {R.}~\bibnamefont {Estevez}}, \bibinfo {author} {\bibfnamefont {L.}~\bibnamefont {Vignoud}}, \ and\ \bibinfo {author} {\bibfnamefont {C.}~\bibnamefont {Olagnon}},\ }\bibfield  {title} {\enquote {\bibinfo {title} {Determination of the {{Young}}'s {{Modulus}} of a {{TiN Thin Film}} by {{Nanoindentation}}: {{Analytical Models}} and {{FEM Simulation}}},}\ }\href {\doibase 10.1380/ejssnt.2012.624} {\bibfield  {journal} {\bibinfo  {journal} {e-Journal of Surface Science and Nanotechnology}\ }\textbf {\bibinfo {volume} {10}},\ \bibinfo {pages} {624--629} (\bibinfo {year} {2012})}\BibitemShut {NoStop}%
\bibitem [{\citenamefont {Sun}, \citenamefont {Bloyce},\ and\ \citenamefont {Bell}(1995)}]{sunFiniteElementAnalysis1995}%
  \BibitemOpen
  \bibfield  {author} {\bibinfo {author} {\bibfnamefont {Y.}~\bibnamefont {Sun}}, \bibinfo {author} {\bibfnamefont {A.}~\bibnamefont {Bloyce}}, \ and\ \bibinfo {author} {\bibfnamefont {T.}~\bibnamefont {Bell}},\ }\bibfield  {title} {\enquote {\bibinfo {title} {Finite element analysis of plastic deformation of various {{TiN}} coating/ substrate systems under normal contact with a rigid sphere},}\ }\href {\doibase 10.1016/0040-6090(95)06942-9} {\bibfield  {journal} {\bibinfo  {journal} {Thin Solid Films}\ }\textbf {\bibinfo {volume} {271}},\ \bibinfo {pages} {122--131} (\bibinfo {year} {1995})}\BibitemShut {NoStop}%
\bibitem [{\citenamefont {Nielsen}\ \emph {et~al.}(2017)\citenamefont {Nielsen}, \citenamefont {Tsaturyan}, \citenamefont {M{\o}ller}, \citenamefont {Polzik},\ and\ \citenamefont {Schliesser}}]{nielsen2017multimode}%
  \BibitemOpen
  \bibfield  {author} {\bibinfo {author} {\bibfnamefont {W.~H.~P.}\ \bibnamefont {Nielsen}}, \bibinfo {author} {\bibfnamefont {Y.}~\bibnamefont {Tsaturyan}}, \bibinfo {author} {\bibfnamefont {C.~B.}\ \bibnamefont {M{\o}ller}}, \bibinfo {author} {\bibfnamefont {E.~S.}\ \bibnamefont {Polzik}}, \ and\ \bibinfo {author} {\bibfnamefont {A.}~\bibnamefont {Schliesser}},\ }\bibfield  {title} {\enquote {\bibinfo {title} {Multimode optomechanical system in the quantum regime},}\ }\href@noop {} {\bibfield  {journal} {\bibinfo  {journal} {Proceedings of the National Academy of Sciences}\ }\textbf {\bibinfo {volume} {114}},\ \bibinfo {pages} {62--66} (\bibinfo {year} {2017})}\BibitemShut {NoStop}%
\bibitem [{\citenamefont {Mohanty}\ \emph {et~al.}(2002)\citenamefont {Mohanty}, \citenamefont {Harrington}, \citenamefont {Ekinci}, \citenamefont {Yang}, \citenamefont {Murphy},\ and\ \citenamefont {Roukes}}]{mohantyIntrinsicDissipationHighfrequency2002a}%
  \BibitemOpen
  \bibfield  {author} {\bibinfo {author} {\bibfnamefont {P.}~\bibnamefont {Mohanty}}, \bibinfo {author} {\bibfnamefont {D.~A.}\ \bibnamefont {Harrington}}, \bibinfo {author} {\bibfnamefont {K.~L.}\ \bibnamefont {Ekinci}}, \bibinfo {author} {\bibfnamefont {Y.~T.}\ \bibnamefont {Yang}}, \bibinfo {author} {\bibfnamefont {M.~J.}\ \bibnamefont {Murphy}}, \ and\ \bibinfo {author} {\bibfnamefont {M.~L.}\ \bibnamefont {Roukes}},\ }\bibfield  {title} {\enquote {\bibinfo {title} {Intrinsic dissipation in high-frequency micromechanical resonators},}\ }\href {\doibase 10.1103/PhysRevB.66.085416} {\bibfield  {journal} {\bibinfo  {journal} {Physical Review B}\ }\textbf {\bibinfo {volume} {66}},\ \bibinfo {pages} {085416} (\bibinfo {year} {2002})}\BibitemShut {NoStop}%
\bibitem [{\citenamefont {Hoehne}\ \emph {et~al.}(2010)\citenamefont {Hoehne}, \citenamefont {Pashkin}, \citenamefont {Astafiev}, \citenamefont {Faoro}, \citenamefont {Ioffe}, \citenamefont {Nakamura},\ and\ \citenamefont {Tsai}}]{hoehneDampingHighfrequencyMetallic2010}%
  \BibitemOpen
  \bibfield  {author} {\bibinfo {author} {\bibfnamefont {F.}~\bibnamefont {Hoehne}}, \bibinfo {author} {\bibfnamefont {{\relax Yu}.~A.}\ \bibnamefont {Pashkin}}, \bibinfo {author} {\bibfnamefont {O.}~\bibnamefont {Astafiev}}, \bibinfo {author} {\bibfnamefont {L.}~\bibnamefont {Faoro}}, \bibinfo {author} {\bibfnamefont {L.~B.}\ \bibnamefont {Ioffe}}, \bibinfo {author} {\bibfnamefont {Y.}~\bibnamefont {Nakamura}}, \ and\ \bibinfo {author} {\bibfnamefont {J.~S.}\ \bibnamefont {Tsai}},\ }\bibfield  {title} {\enquote {\bibinfo {title} {Damping in high-frequency metallic nanomechanical resonators},}\ }\href {\doibase 10.1103/PhysRevB.81.184112} {\bibfield  {journal} {\bibinfo  {journal} {Physical Review B}\ }\textbf {\bibinfo {volume} {81}},\ \bibinfo {pages} {184112} (\bibinfo {year} {2010})}\BibitemShut {NoStop}%
\bibitem [{\citenamefont {Hauer}\ \emph {et~al.}(2018)\citenamefont {Hauer}, \citenamefont {Kim}, \citenamefont {Doolin}, \citenamefont {Souris},\ and\ \citenamefont {Davis}}]{hauerTwolevelSystemDamping2018a}%
  \BibitemOpen
  \bibfield  {author} {\bibinfo {author} {\bibfnamefont {B.~D.}\ \bibnamefont {Hauer}}, \bibinfo {author} {\bibfnamefont {P.~H.}\ \bibnamefont {Kim}}, \bibinfo {author} {\bibfnamefont {C.}~\bibnamefont {Doolin}}, \bibinfo {author} {\bibfnamefont {F.}~\bibnamefont {Souris}}, \ and\ \bibinfo {author} {\bibfnamefont {J.~P.}\ \bibnamefont {Davis}},\ }\bibfield  {title} {\enquote {\bibinfo {title} {Two-level system damping in a quasi-one-dimensional optomechanical resonator},}\ }\href {\doibase 10.1103/PhysRevB.98.214303} {\bibfield  {journal} {\bibinfo  {journal} {Physical Review B}\ }\textbf {\bibinfo {volume} {98}},\ \bibinfo {pages} {214303} (\bibinfo {year} {2018})}\BibitemShut {NoStop}%
\bibitem [{\citenamefont {Zhou}\ \emph {et~al.}(2019)\citenamefont {Zhou}, \citenamefont {Cattiaux}, \citenamefont {Gazizulin}, \citenamefont {Luck}, \citenamefont {Maillet}, \citenamefont {Crozes}, \citenamefont {Motte}, \citenamefont {Bourgeois}, \citenamefont {Fefferman},\ and\ \citenamefont {Collin}}]{zhouOnchipThermometryMicrowave2019}%
  \BibitemOpen
  \bibfield  {author} {\bibinfo {author} {\bibfnamefont {X.}~\bibnamefont {Zhou}}, \bibinfo {author} {\bibfnamefont {D.}~\bibnamefont {Cattiaux}}, \bibinfo {author} {\bibfnamefont {R.~R.}\ \bibnamefont {Gazizulin}}, \bibinfo {author} {\bibfnamefont {A.}~\bibnamefont {Luck}}, \bibinfo {author} {\bibfnamefont {O.}~\bibnamefont {Maillet}}, \bibinfo {author} {\bibfnamefont {T.}~\bibnamefont {Crozes}}, \bibinfo {author} {\bibfnamefont {J.-F.}\ \bibnamefont {Motte}}, \bibinfo {author} {\bibfnamefont {O.}~\bibnamefont {Bourgeois}}, \bibinfo {author} {\bibfnamefont {A.}~\bibnamefont {Fefferman}}, \ and\ \bibinfo {author} {\bibfnamefont {E.}~\bibnamefont {Collin}},\ }\bibfield  {title} {\enquote {\bibinfo {title} {On-chip {{Thermometry}} for {{Microwave Optomechanics Implemented}} in a {{Nuclear Demagnetization Cryostat}}},}\ }\href {\doibase 10.1103/PhysRevApplied.12.044066} {\bibfield  {journal} {\bibinfo  {journal} {Physical Review Applied}\ }\textbf {\bibinfo {volume} {12}},\ \bibinfo {pages} {044066} (\bibinfo
  {year} {2019})}\BibitemShut {NoStop}%
\end{thebibliography}%


\begin{thebibliography}{4}%
\makeatletter
\providecommand \@ifxundefined [1]{%
 \@ifx{#1\undefined}
}%
\providecommand \@ifnum [1]{%
 \ifnum #1\expandafter \@firstoftwo
 \else \expandafter \@secondoftwo
 \fi
}%
\providecommand \@ifx [1]{%
 \ifx #1\expandafter \@firstoftwo
 \else \expandafter \@secondoftwo
 \fi
}%
\providecommand \natexlab [1]{#1}%
\providecommand \enquote  [1]{``#1''}%
\providecommand \bibnamefont  [1]{#1}%
\providecommand \bibfnamefont [1]{#1}%
\providecommand \citenamefont [1]{#1}%
\providecommand \href@noop [0]{\@secondoftwo}%
\providecommand \href [0]{\begingroup \@sanitize@url \@href}%
\providecommand \@href[1]{\@@startlink{#1}\@@href}%
\providecommand \@@href[1]{\endgroup#1\@@endlink}%
\providecommand \@sanitize@url [0]{\catcode `\\12\catcode `\$12\catcode `\&12\catcode `\#12\catcode `\^12\catcode `\_12\catcode `\%12\relax}%
\providecommand \@@startlink[1]{}%
\providecommand \@@endlink[0]{}%
\providecommand \url  [0]{\begingroup\@sanitize@url \@url }%
\providecommand \@url [1]{\endgroup\@href {#1}{\urlprefix }}%
\providecommand \urlprefix  [0]{URL }%
\providecommand \Eprint [0]{\href }%
\providecommand \doibase [0]{http://dx.doi.org/}%
\providecommand \selectlanguage [0]{\@gobble}%
\providecommand \bibinfo  [0]{\@secondoftwo}%
\providecommand \bibfield  [0]{\@secondoftwo}%
\providecommand \translation [1]{[#1]}%
\providecommand \BibitemOpen [0]{}%
\providecommand \bibitemStop [0]{}%
\providecommand \bibitemNoStop [0]{.\EOS\space}%
\providecommand \EOS [0]{\spacefactor3000\relax}%
\providecommand \BibitemShut  [1]{\csname bibitem#1\endcsname}%
\let\auto@bib@innerbib\@empty
\bibitem [{\citenamefont {Sun}\ \emph {et~al.}(2015)\citenamefont {Sun}, \citenamefont {Makise}, \citenamefont {Qiu}, \citenamefont {Terai},\ and\ \citenamefont {Wang}}]{sunFabrication200Oriented2015}%
  \BibitemOpen
  \bibfield  {author} {\bibinfo {author} {\bibfnamefont {R.}~\bibnamefont {Sun}}, \bibinfo {author} {\bibfnamefont {K.}~\bibnamefont {Makise}}, \bibinfo {author} {\bibfnamefont {W.}~\bibnamefont {Qiu}}, \bibinfo {author} {\bibfnamefont {H.}~\bibnamefont {Terai}}, \ and\ \bibinfo {author} {\bibfnamefont {Z.}~\bibnamefont {Wang}},\ }\href {\doibase 10.1109/TASC.2014.2383694} {\bibfield  {journal} {\bibinfo  {journal} {IEEE Transactions on Applied Superconductivity}\ }\textbf {\bibinfo {volume} {25}},\ \bibinfo {pages} {1} (\bibinfo {year} {2015})}\BibitemShut {NoStop}%
\bibitem [{\citenamefont {Tsaturyan}\ \emph {et~al.}(2014)\citenamefont {Tsaturyan}, \citenamefont {Barg}, \citenamefont {Simonsen}, \citenamefont {Villanueva}, \citenamefont {Schmid}, \citenamefont {Schliesser},\ and\ \citenamefont {Polzik}}]{tsaturyanDemonstrationSuppressedPhonon2014}%
  \BibitemOpen
  \bibfield  {author} {\bibinfo {author} {\bibfnamefont {Y.}~\bibnamefont {Tsaturyan}}, \bibinfo {author} {\bibfnamefont {A.}~\bibnamefont {Barg}}, \bibinfo {author} {\bibfnamefont {A.}~\bibnamefont {Simonsen}}, \bibinfo {author} {\bibfnamefont {L.~G.}\ \bibnamefont {Villanueva}}, \bibinfo {author} {\bibfnamefont {S.}~\bibnamefont {Schmid}}, \bibinfo {author} {\bibfnamefont {A.}~\bibnamefont {Schliesser}}, \ and\ \bibinfo {author} {\bibfnamefont {E.~S.}\ \bibnamefont {Polzik}},\ }\href {\doibase 10.1364/OE.22.006810} {\bibfield  {journal} {\bibinfo  {journal} {Optics Express}\ }\textbf {\bibinfo {volume} {22}},\ \bibinfo {pages} {6810} (\bibinfo {year} {2014})}\BibitemShut {NoStop}%
\bibitem [{\citenamefont {Yu}\ \emph {et~al.}(2014)\citenamefont {Yu}, \citenamefont {Cicak}, \citenamefont {Kampel}, \citenamefont {Tsaturyan}, \citenamefont {Purdy}, \citenamefont {Simmonds},\ and\ \citenamefont {Regal}}]{yuPhononicBandgapShield2014}%
  \BibitemOpen
  \bibfield  {author} {\bibinfo {author} {\bibfnamefont {P.-L.}\ \bibnamefont {Yu}}, \bibinfo {author} {\bibfnamefont {K.}~\bibnamefont {Cicak}}, \bibinfo {author} {\bibfnamefont {N.~S.}\ \bibnamefont {Kampel}}, \bibinfo {author} {\bibfnamefont {Y.}~\bibnamefont {Tsaturyan}}, \bibinfo {author} {\bibfnamefont {T.~P.}\ \bibnamefont {Purdy}}, \bibinfo {author} {\bibfnamefont {R.~W.}\ \bibnamefont {Simmonds}}, \ and\ \bibinfo {author} {\bibfnamefont {C.~A.}\ \bibnamefont {Regal}},\ }\href {\doibase 10.1063/1.4862031} {\bibfield  {journal} {\bibinfo  {journal} {Applied Physics Letters}\ }\textbf {\bibinfo {volume} {104}},\ \bibinfo {pages} {023510} (\bibinfo {year} {2014})}\BibitemShut {NoStop}%
\bibitem [{\citenamefont {Beccari}\ \emph {et~al.}(2022)\citenamefont {Beccari}, \citenamefont {Visani}, \citenamefont {Fedorov}, \citenamefont {Bereyhi}, \citenamefont {Boureau}, \citenamefont {Engelsen},\ and\ \citenamefont {Kippenberg}}]{beccariStrainedCrystallineNanomechanical2022}%
  \BibitemOpen
  \bibfield  {author} {\bibinfo {author} {\bibfnamefont {A.}~\bibnamefont {Beccari}}, \bibinfo {author} {\bibfnamefont {D.~A.}\ \bibnamefont {Visani}}, \bibinfo {author} {\bibfnamefont {S.~A.}\ \bibnamefont {Fedorov}}, \bibinfo {author} {\bibfnamefont {M.~J.}\ \bibnamefont {Bereyhi}}, \bibinfo {author} {\bibfnamefont {V.}~\bibnamefont {Boureau}}, \bibinfo {author} {\bibfnamefont {N.~J.}\ \bibnamefont {Engelsen}}, \ and\ \bibinfo {author} {\bibfnamefont {T.~J.}\ \bibnamefont {Kippenberg}},\ }\href {\doibase 10.1038/s41567-021-01498-4} {\bibfield  {journal} {\bibinfo  {journal} {Nature Physics}\ }\textbf {\bibinfo {volume} {18}},\ \bibinfo {pages} {436} (\bibinfo {year} {2022})}\BibitemShut {NoStop}%
\end{thebibliography}%

\end{document}


\title{Supplementary Material: High-$Q$ membrane resonators using ultra-high stressed crystalline TiN films}

\author{Yuki Matsuyama}
\affiliation{Komaba Institute for Science (KIS), The University of Tokyo, Meguro-ku, Tokyo 153-8902, Japan}

\author{Shotaro Shirai}
\affiliation{RIKEN Center for Quantum Computing (RQC), Wako, Saitama 351–0198, Japan}
\affiliation{Komaba Institute for Science (KIS), The University of Tokyo, Meguro-ku, Tokyo 153-8902, Japan}

\author{Ippei Nakamura}
\affiliation{Komaba Institute for Science (KIS), The University of Tokyo, Meguro-ku, Tokyo 153-8902, Japan}

\author{Masao Tokunari}
\affiliation{IBM Quantum, IBM Research - Tokyo,
Chuo-ku, Tokyo, 103-8510, Japan}

\author{Hirotaka Terai}
\affiliation{Advanced ICT Research
Institute, National Institute of Information and Communications Technology,
Kobe, Hyogo, 651-2492, Japan}

\author{Yuji Hishida}
\affiliation{Advanced ICT Research
Institute, National Institute of Information and Communications Technology,
Kobe, Hyogo, 651-2492, Japan}

\author{Ryo Sasaki}
\affiliation{RIKEN Center for Quantum Computing (RQC), Wako, Saitama 351–0198, Japan}

\author{Yusuke Tominaga}
\affiliation{RIKEN Center for Quantum Computing (RQC), Wako, Saitama 351–0198, Japan}

\author{Atsushi Noguchi}
\affiliation{Komaba Institute for Science (KIS), The University of Tokyo, Meguro-ku, Tokyo 153-8902, Japan}
\affiliation{RIKEN Center for Quantum Computing (RQC), Wako, Saitama 351–0198, Japan}
\affiliation{Inamori Research Institute for Science (InaRIS), Kyoto-shi, Kyoto 600-8411, Japan}

\date{\today}

\maketitle

\section{Fabrication process}

Here, we briefly describe the fabrication process for TiN membrane resonators with phononic crystals. We started with a 3-inch, 300-$\mathrm{\mu m}$-thick, high resistivity intrinsic silicon substrate with a 100-nm-thick layer of titanium nitride (TiN), which was deposited at 880 $^\circ$C by DC magnetron sputtering\cite{sunFabrication200Oriented2015}.
After the TiN was patterned by reactive ion etching (RIE), the photoresist layer was removed, followed by cleaning of its surface with oxygen plasma ashing.
Silicon dioxide ($\mathrm{SiO_2}$) was deposited using chemical vapor depositon (CVD) on the TiN side to protect the TiN surface and on the other side (Si side) to be patterned as a mask for Si etching.
The $\mathrm{SiO_2}$ layer on the Si side was patterned by RIE, followed by Si etching using deep reactive ion etching (DRIE).
The DRIE process was stopped before fully etching through the Si substrate, and the remaining Si was etched by wet etching with tetramethylammonium hydroxide (TMAH).
The wafer was diced into chips after the DRIE process, and wet etching was performed for each chip.
After completion of the Si etching, the $\mathrm{SiO_2}$ layers were removed with hydrofluoric (HF) acid, and finally the TiN membrane structure was released.

\begin{figure}
\centering
\includegraphics[width=5.0in]{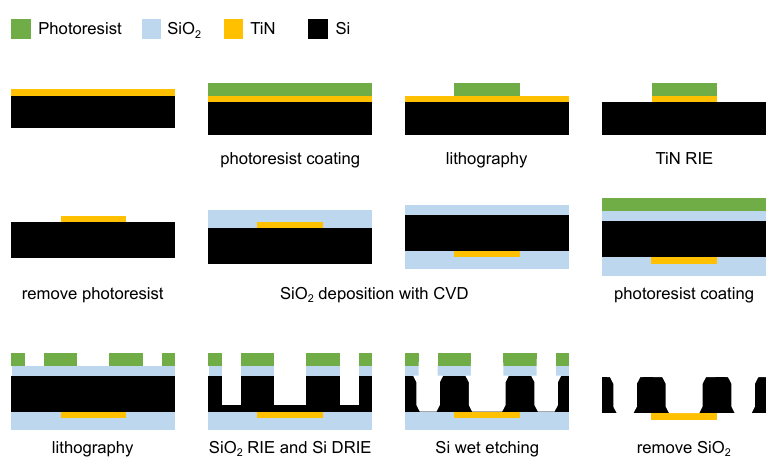}
\caption{An illustration of fabrication process for TiN membrane resonators with Si phononic crystals.}
\label{fig:S1}
\end{figure}

\section{Designs of phononic crystals}
\begin{figure}
\centering
\includegraphics[width=3.37in]{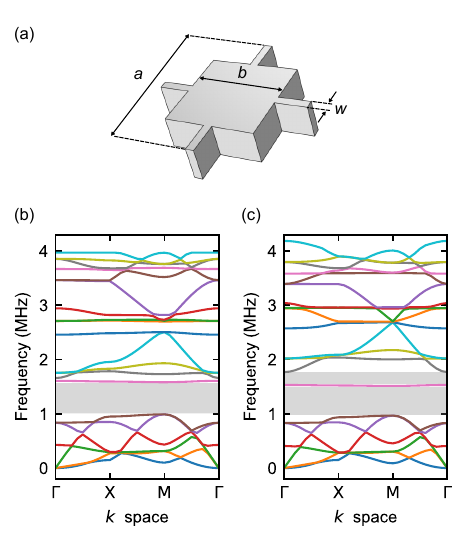}
\caption{Simulated diagrams of phononic bands. (a) Schematic of the unit cell of our phononic crystal. (b) Phononic band diagram calculated using the actual design parameters: $a=\mathrm{1500\,\mu m},\,b= \mathrm{900\,\mu m},\,w= \mathrm{90\,\mu m}$. (c) Phononic band diagram calculated using the parameters slightly detuned from (b): $a=\mathrm{1450\,\mu m},\,b= \mathrm{925\,\mu m},\,w= \mathrm{80\,\mu m}$.}
\label{fig:S2}
\end{figure}

We designed the phononic crystal following the geometries proposed in previous works\cite{tsaturyanDemonstrationSuppressedPhonon2014,yuPhononicBandgapShield2014}.
Finite-element-method (FEM) simulations with COMSOL Multiphysics were performed to determine the design parameters of the phononic crystal. 
Phononic band structures were calculated by assuming an infinite array of unit cells with periodic boundary conditions.
The unit cell and swept parameters are shown in Fig. \ref{fig:S2}(a).

Figure \ref{fig:S2}(b) represents the phononic band structure calculated using the design parameters of the actual device.
A single bandgap appears around 1.0--1.6 MHz.
This result does not include the 1.65--1.85 MHz phononic bandgap observed in the experimental results (Fig. 2(a) in the main text).
We attribute this discrepancy to slight deviations of the parameters, caused by side etching during the wet etching process. 
Indeed, when the parameters are slightly adjusted in the FEM simulation, the second bandgap appears around 1.65--1.85 MHz, as shown in Fig. \ref{fig:S2}(c).

\section{Measurement setup}

\begin{figure}
\centering
\includegraphics[width=5.0in]{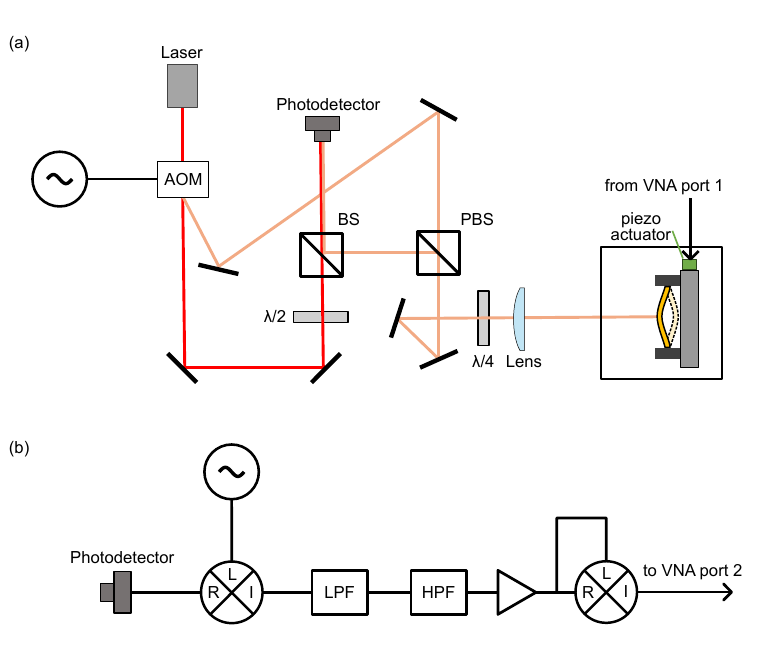}
\caption{Schematic of the heterodyne experimental setup. (a) Schematic of the Michelson interferometer setup. The laser of $\lambda = 626$ nm is used for the measurement, which is shown as red path.
The frequency of the laser which is shed on the membrane resonator is shifted by acousto-optic mudulator.
To aid visualization, the frequency-shifted light is depicted in light red. (b) Schematic of the RF filtering setup.}
\label{fig:S3}
\end{figure}

We measured the oscillation of the TiN membrane resonator using the Michelson interferometer (Fig. \ref{fig:S3}(a)).
The membrane resonator was placed inside a cryostat, which introduced vibrational noise into the measurement.
We employed heterodyne detection to shift the interference signal to the RF band and to separate the carrier and sidebands.
Noise cancellation was achieved by focusing on the three frequency components of the laser light modulated by the membrane resonator: carrier ($\omega_L$), upper sideband ($\omega_L + \omega_m$), and lower sideband ($\omega_L - \omega_m$). 
As all three signals are subject to the same vibration noise, subtracting two of them results in its removal.

The optical measurement setup is shown in Fig. \ref{fig:S3}(a).
A 626-nm laser was employed in the measurement. 
Its frequency was shifted by 80 MHz using an acousto-optic modulator (AOM), and the frequency-shifted beam was focused onto the membrane resonator. 
The reflected beam then interfered with the reference beam, and the resulting interference signal was detected by a photodetector.
To suppress heating of the membrane resonator, the power of the frequency-shifted beam was kept as low as possible.
The signal frequency was down-converted from the laser frequency ($\omega_L$) to the AOM frequency ($\omega_\mathrm{AOM}$)  by interference.

Figure \ref{fig:S3} (b) shows the RF filtering setup used to cancel out the noise.
The signal detected by the photodetector was mixed with a local oscillator signal of a frequency $\omega_\mathrm{AOM}-\omega_a$, which is slightly lower than the AOM frequency. For instance, a signal with a frequency of 78.6 MHz was used for measurement of the fundamental mode at 1.132 MHz.
A low-pass filter with a cutoff frequency of 140 MHz was used to extract the down-converted signal, and then the carrier, upper sideband, and lower sideband frequencies are denoted as $\omega_a$, $\omega_m+\omega_a$, and $\omega_m -\omega_a$, respectively.
The lower sideband signal was then eliminated by a high-pass filter with a cutoff frequency of 230 kHz.
Finally, the carrier and upper sideband signals were mixed, yielding a noise-reduced signal from the membrane resonator.
A vector network analyzer (VNA) was used to drive and measure the oscillation of the membrane resonator.

\section{Resonant frequency shifts in low temperautre measurements}

\begin{figure}
\centering
\includegraphics[width=0.8\textwidth]{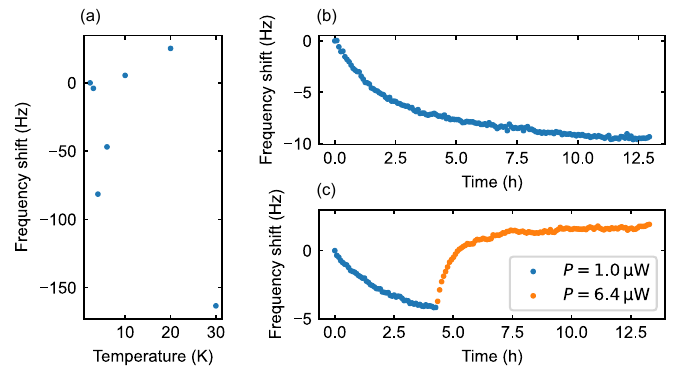}
\caption{Resonant frequency shifts observed during low temperature measurements of the fundamental mode. (a) The temperature dependence below 30 K. (b) The resonant frequency shift after stopping the vacuum pumping. (c) The resonant frequency shift induced by changing the laser power $P$. The measurement was performed right after $P$ was changed from 6.4 $\mathrm{\mu W}$ to 1.0 $\mathrm{\mu W}$} 
\label{fig:S4}
\end{figure}

Here, we report the resonant frequency shift observed in low temperature measurements (Fig. \ref{fig:S4}). 
Since resonant frequencies of a square membrane resonator are determined by its geometry, tensile stress, and density of the material, any frequency variation of a single membrane resonator is expected to originate from the tensile stress changes.
However, we observed small resonant frequency shifts that could not be attributed to the tensile stress variation.

Figure \ref{fig:S4} (a) shows the fundamental mode frequency shift observed below 30 K during the temperature dependence experiment (Fig. 3 in the main text).
From 2.2 K, the temperature was increased and controlled by the temperature controller.
Although the resonant frequency is expected to decrease with increasing temperature due to the reduction of the tensile stress, it exhibited a non-monotonic behavior. The resonant frequency was shifted even when the temperature was kept constant. 
Figure \ref{fig:S4}(b) represents the time evolution of the fundamental mode frequency at the base temperature after turning off the vacuum pump.
The resonant frequency decreased over time and continued to drop even after 10 hours.

Changes in the power of the measurement laser also induced the resonant frequency shifts (Fig. \ref{fig:S4} (c)).
The resonant frequency increased at high power and decreased at low power, which is opposite to the heating effect that reduces the frequency by lowering the tensile stress.
Although only the fundamental mode is shown here, similar frequency shifts were also observed in higher-order modes.

From these results, we attribute the resonant frequency shifts to changes in the mean density of the membrane due to adsorption and desorption of molecules on the membrane surface.
We have not yet found a model that explains the frequency shifts occurring on a timescale of several hours.
Previous work reported a reversed increase of the mechanical dissipation around 10 K, which was attributed to slow condensation of residual gases on the sample\cite{beccariStrainedCrystallineNanomechanical2022}.

\bibliography{supplementary}

